\documentclass[sigconf]{acmart}

\AtBeginDocument{%
  }

\copyrightyear{2026}
\acmYear{2026}
\setcopyright{cc}
\setcctype{by}
\acmConference[UIST '26]{The 39th Annual ACM Symposium on User Interface Software and Technology}{November 02--05, 2026}{Detroit, MI, USA}
\acmBooktitle{The 39th Annual ACM Symposium on User Interface Software and Technology (UIST '26), November 02--05, 2026, Detroit, MI, USA}
\acmDOI{10.1145/3830398.3830637}
\acmISBN{979-8-4007-2856-3/2026/11}

\usepackage{framed}
\usepackage{xspace}
\usepackage{enumitem}

\definecolor{revisioncolor}{RGB}{0, 0, 0}

\begin{document}

\title{Penquiry: A Pen-based Interactive In-situ Q\&A System Leveraging LLMs}

\author{Jeongmin Rhee}
\authornote{Both authors contributed equally to this research.}
\affiliation{%
  \institution{Seoul National University}
  \city{Seoul}
  \country{Republic of Korea}
}
\email{jmrhee@hcil.snu.ac.kr}
\orcid{0009-0007-3533-6603}

\author{Changhee Lee}
\authornotemark[1]
\affiliation{%
  \institution{Pohang University\\of Science and Technology}
  \city{Pohang}
  \country{Republic of Korea}
}
\email{chlee0811@postech.ac.kr}
\orcid{0009-0006-2392-1285}

\author{Hyunwoo Kim}
\affiliation{%
  \institution{Seoul National University}
  \city{Seoul}
  \country{Republic of Korea}
}
\email{hyunkim@hcil.snu.ac.kr}
\orcid{0009-0007-1827-165X}

\author{Kiroong Choe}
\affiliation{%
  \institution{Seoul National University}
  \city{Seoul}
  \country{Republic of Korea}
}
\email{krchoe@hcil.snu.ac.kr}
\orcid{0000-0002-6084-2539}

\author{Bohyoung Kim}
\authornotemark[2]
\affiliation{%
  \institution{Hankuk University\\of Foreign Studies}
  \city{Yongin-si}
  \country{Republic of Korea}
}
\email{bkim@hufs.ac.kr}
\orcid{0000-0002-2183-5651}

\author{Sungahn Ko}
\authornotemark[2]
\affiliation{%
  \institution{Pohang University\\of Science and Technology}
  \city{Pohang}
  \country{Republic of Korea}
}
\email{sungahn@postech.ac.kr}
\orcid{0000-0002-7410-5652}

\author{Jinwook Seo}
\authornote{Corresponding Author.}
\affiliation{%
  \institution{Seoul National University}
  \city{Seoul}
  \country{Republic of Korea}
}
\email{jseo@snu.ac.kr}
\orcid{0000-0002-7734-822X}

\renewcommand{\shortauthors}{Rhee, Lee, et al.}

\begin{abstract}
Pen-based digital devices remain a preferred medium for active, cognitively engaging study. Concurrently, Large Language Models (LLMs) have become indispensable for self-directed learning, enabling students to clarify concepts. However, a fundamental interaction gap exists between the fluid, spatial nature of pen-based workflows and the discrete, keyboard-heavy requirements of LLMs. We present \penquiry, an in-situ question-and-answer system that bridges this gap by enabling learners to pose questions directly on digital study materials via a pen. We characterize two primary interaction challenges in this multimodal transition: a Referential Barrier, which hinders \textcolor{revisioncolor}{grounding fine-grained visual elements into the query context}, and an Expressive Barrier, which forces learners to translate diverse, non-textual intents---such as equations and diagrams---into rigid, typed sentences. To resolve these, \penquiry~introduces a mediation layer featuring \contentsnapping~ for unambiguous referencing and \questionautocompletion~ to expand sparse ink keywords into rich semantic queries. Through two iterative user studies ($N=16$ per study), we demonstrate that \penquiry~significantly reduces the cognitive and physical overhead of inquiry compared to traditional interfaces, providing a new blueprint for pen-based, in-situ AI interaction.
\end{abstract}

\ccsdesc[500]{Human-centered computing~User interface programming}
\ccsdesc[300]{Human-centered computing~Interaction techniques}
\ccsdesc[100]{Human-centered computing~Touch screens}

\keywords{In-situ Interaction, Pen-based Interaction, Large Language Models, Pen-based Learning, Human-AI Interaction}
\begin{teaserfigure}
    \centering
    \includegraphics[width=\linewidth]{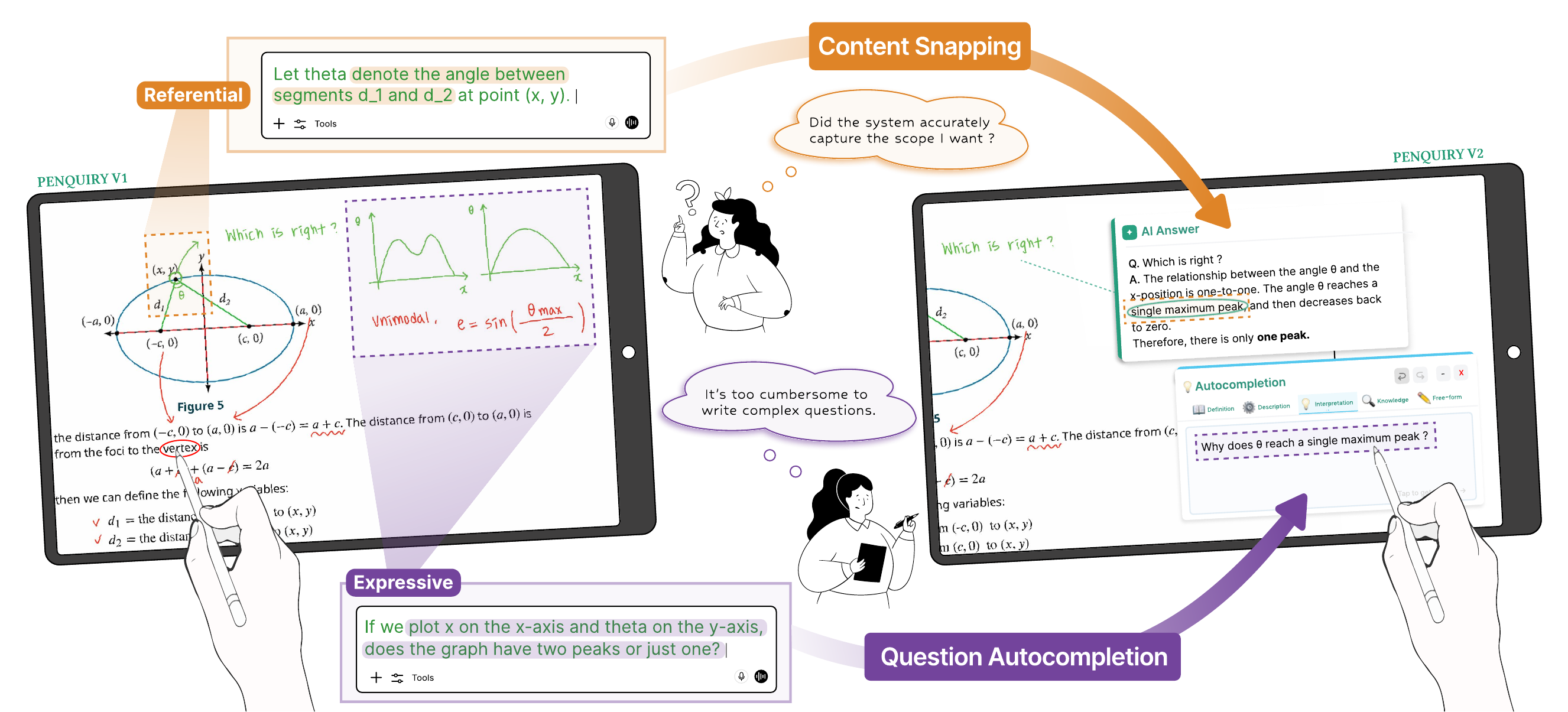}
    \caption{\penquiry~is an in-situ Q\&A system that enables learners to query directly on digital study materials using a pen. (Left) \penquiry~V1 leverages in-situ pen input to mitigate the referential and expressive barriers inherent in a separate keyboard interface for LLM queries. However, it introduces new challenges, such as uncertainty in referential grounding and the physical burden of handwriting lengthy questions. (Right) \penquiry~V2 resolves these emergent barriers by introducing \contentsnapping, which \textcolor{revisioncolor}{aligns user sketches with nearby document text elements}, and \questionautocompletion, which generates full queries from minimal ink keywords. \textcolor{revisioncolor}{Together, these mechanisms mediate free-form pen input into LLM-interpretable questions while sustaining inquiry within the learner's pen-centered workflow.}}
    \label{fig:teaser}
\end{teaserfigure}


\definecolor{TableauBlue}{HTML}{1F77B4}
\definecolor{TableauOrange}{HTML}{FF7F0E}
\definecolor{TableauGreen}{HTML}{2CA02C}
\definecolor{TableauRed}{HTML}{D62728}
\definecolor{TableauPurple}{HTML}{9467BD}
\definecolor{TableauBrown}{HTML}{8C564B}
\definecolor{TableauPink}{HTML}{E377C2}
\definecolor{TableauGray}{HTML}{7F7F7F}
\definecolor{TableauOlive}{HTML}{BCBD22}
\definecolor{TableauCyan}{HTML}{17BECF}

\newcommand{\penquiry}{\textsc{Pen\-q{}ui\-ry}}

\newcommand{\barrierbox}[2]{%
  \tcbox[on line,
    colback=#1!30,
    boxrule=0pt,
    arc=3pt,
    left=0.5pt, right=0.5pt, top=0.5pt, bottom=0.5pt]{#2}
}

\newcommand{\RB}{{Referential Barrier}}
\newcommand{\EB}{{Expressive Barrier}}

\definecolor{diffbg}{RGB}{255, 255, 255}
\definecolor{difftxt}{RGB}{0, 0, 0}
\definecolor{nondiffbg}{RGB}{255, 255, 255}
\definecolor{nondifftxt}{RGB}{0,0,0}
\newcommand{\diff}[1]{\textcolor{difftxt}{{#1}}}
\definecolor{revisedcolor}{RGB}{127, 17, 16}

\newcommand{\contentsnapping}{{Content Snapping}}
\newcommand{\questionautocompletion}{{Question Autocompletion}}

\maketitle

\section{Introduction}
\label{sec:Introduction}

Pen-based digital environments remain a preferred method for active learning due to their cognitive benefits \cite{mueller2014pen, ihara2021advantage, oviatt2010toward, shell2021make}. Simultaneously, learners increasingly rely on Large Language Models (LLMs) as indispensable tools for real-time concept clarification \cite{sharma2025role, steinert2023harnessing} during self-directed study. However, integrating these two paradigms reveals a fundamental interaction gap: the fluid and spatial nature of stylus workflows is at odds with the discrete text-heavy requirements of current LLM interfaces. Because querying an LLM typically requires abandoning the document canvas for a separate, keyboard-driven chat window, learners face a severe modality mismatch. This constant context-switching not only fractures the learner's focus but also imposes a high interaction cost, often discouraging inquiry and disrupting the flow of learning \cite{holstein2019designing}.

\textcolor{revisioncolor}{While in-situ multimodal approaches combining pen and speech \newline~\cite{huang2025sketchgpt, 10.1145/258549.258821} can reduce such constant context-switching, speech input remains limited by situational constraints, such as quiet libraries or noisy public spaces, and lacks the fine-grained spatial precision of stylus input. Therefore, we deliberately scope this work to pen-only interaction, seeking to understand how LLM-based inquiry can be fully realized through direct, ink-based modalities without disrupting existing learning workflows.}

To empirically understand the friction caused by this modality mismatch, we conducted a formative study \textcolor{revisioncolor}{($N=6$)} observing learners navigating standard, out-of-situ LLM workflows. Our findings reveal that learners struggle to map their spatial and visual intent into the discrete text required by the system. We characterize this friction as two distinct interaction challenges: (1) a \RB, the difficulty of precisely grounding inquiries to specific visual regions; and (2) an \EB, the high physical and cognitive overhead of articulating non-textual concepts via standard keyboard input.

These barriers manifest as bottlenecks in the user's workflow. The \RB~ arises because learners must provide \textcolor{revisioncolor}{detailed} textual descriptions for visual elements. For example, to simply reference an angle $\theta$ in a diagram (Figure 1, Top Left), a learner must construct a lengthy, unambiguous prompt (e.g., ``Let theta denote the angle between segments $d_1$ and $d_2$ at point $(x, y)$''), making the mere act of referencing inefficient. Similarly, the \EB~ stems from the low bandwidth of text-only input when dealing with symbolic or structural information. When questioning the shape of a mathematical graph (Figure 1, Bottom Left), learners are forced to translate 2D visual properties into cumbersome 1D text strings (e.g., ``If we plot $x$ on the horizontal axis and $\theta$ on the vertical axis, does the graph have two peaks or just one?'').

To address these barriers, we designed and implemented \textbf{\penquiry~Version 1} (V1), an in-situ system enabling learners to query directly on study materials via digital ink. In our first user study ($N=16$), a comparison against a baseline ex-situ keyboard environment demonstrated that V1 mitigated initial barriers by allowing users to directly point to visual elements and freely sketch formulas. However, this study revealed that simply shifting the interface to the pen introduced secondary interaction challenges stemming from raw, unmediated ink input. Regarding the \RB, high degrees of freedom inherent to free-form sketches created interpretive ambiguity; without explicit visual grounding, learners could not verify whether the system accurately bounded their intended referential scope. Regarding the \EB, while sketching bypassed the need for textual workarounds, the high physical effort of handwriting lengthy, semantically complex prompts severely hampered learners' inquiry bandwidth.

To resolve the friction of unmediated ink, we developed \textbf{\penquiry~Version 2} (V2), which introduces an LLM-mediated layer consisting of two core interaction techniques. First, to eliminate the interpretive ambiguity associated with the \RB, we implemented \contentsnapping, which automatically aligns free-form sketches with document elements, \textcolor{revisioncolor}{building on prior work on imprecise pen selection \cite{lank2005sloppy},} to provide immediate visual confirmation of the system's spatial understanding. Second, to overcome the bandwidth constraints of the \EB, we integrated \questionautocompletion, an LLM-driven feature that synthesizes complete, context-aware candidate questions from sparse ink keywords, reducing the physical overhead of handwriting lengthy queries. In our second user study comparing V1 and V2 ($N=16$), we found that \contentsnapping~ successfully disambiguated the learner's referential intent. Furthermore, \questionautocompletion~ did more than just reduce physical effort; it actively augmented learners' inquiry behavior, scaffolding their thought processes and prompting them to explore complex concepts beyond their initial intent.

Our core contributions are:

\begin{enumerate}
    \item We formalize the \RB~(spatial grounding) and the \EB~(input bandwidth) that characterize the modality mismatch between fluid pen-based workflows and discrete LLM interfaces in pen-based digital learning contexts.

    \item We present the iterative design and implementation of \penquiry. While Version 1 establishes the foundational in-situ questioning framework, Version 2 introduces an intelligent mediation layer featuring \contentsnapping~ and \questionautocompletion~ (to synthesize sparse ink into rich prompts).

    \item Through two user studies ($N=16$ per study), we demonstrate that \penquiry~not only significantly reduces the physical and cognitive overhead of in-situ inquiry, but also actively scaffolds and augments learners' complex questioning behaviors, providing empirical insights for pen-centric AI interaction.

\end{enumerate}

\section{Related Work}
\label{sec:Related_Work}

\subsection{LLMs in Education}

The advent of LLMs has fundamentally invigorated research on conversational learning agents. Recent LLM-based systems facilitate more flexible, natural tutoring dialogues \cite{chen2024empowering, liu2024personality} and have been extensively studied in domains like mathematics to explore both their pedagogical potential and inherent risks \cite{kumar2025math}. Beyond merely providing direct answers, LLM-powered interfaces are increasingly designed to scaffold strategic thinking and complex problem-solving processes \cite{grassucci2025beyond, dinucu2025problem, ogata2024exait}. Furthermore, survey-based research indicates that while generative AI offers transformative opportunities in the classroom, it also necessitates significant pedagogical and institutional adaptation \cite{bower2024should, zawacki2019systematic}.

These advancements represent a paradigm shift toward more adaptive, learner-centered experiences. However, while most existing systems still rely on disjointed, chat-style interfaces, the emergence of multimodal interaction highlights opportunities to embed questioning and feedback more seamlessly into the learning environment \cite{alnajjar2024exploring, kuchemann2025opportunities, kortemeyer2023toward, hinckley2010pen+}. Among these, pen-based interaction stands out as a particularly promising modality, enabling learners to engage in flexible, expressive, and contextually grounded inquiry without the cognitive disruption inherent in traditional text-based inputs.

\subsection{Pen-based Learning}

The cognitive benefits of handwriting and pen-based interaction have been consistently validated in educational and cognitive science research. Compared to typing, handwriting more effectively reinforces memory and conceptual understanding by engaging specific motor and spatial processing mechanisms \cite{marano2025neuroscience, ihara2021advantage, mueller2014pen}. In digital contexts, pen-based systems have similarly been shown to promote metacognitive strategies and improve the overall quality of learning \cite{tan2020effectiveness, al2024effect, van2017only, chimuma2016assessing, Chi02102014}.

Despite these advantages, most contemporary digital note-taking tools treat pen annotations as static marks rather than interactive, functional elements \cite{tashman2011liquidtext}. To address this, prior research has introduced techniques that dynamically adjust document layouts to accommodate ink \cite{yoon2013texttearing, romat2019spaceink} or treat handwriting as an active input that interacts with underlying content structures \cite{romat2019activeink, 10.1145/3025453.3025716}.
Building on these advancements, our work aims to bridge the disconnect between the act of inquiry and the process of information retrieval \cite{weibel2011immersion}. By transforming handwritten annotations into dynamic query inputs, our approach more effectively supports generative learning processes, allowing learners to remain immersed in their primary study workflow.

\subsection{Interaction Paradigms beyond Chat and Keyboard Input}

While conversational interfaces have lowered the barrier to LLM adoption, they inherently decouple the user’s active workspace from the interaction interface. To bridge this gap, recent Human-Computer Interaction (HCI) research has applied Direct Manipulation principles ~\cite{shneiderman1983direct} to AI interactions. By acting directly on document content, these systems \textcolor{revisioncolor}{use text selections, cursor positions, or predefined elements} as implicit prompts, reducing the cognitive effort required to articulate context ~\cite{barke2023grounded, masson2024directgpt, suh2024luminate, suh2023sensecape}. However, \textcolor{revisioncolor}{such approaches are less suited to spatially diffuse regions that lack explicit structural boundaries but remain semantically meaningful to users.}

To move beyond \textcolor{revisioncolor}{keyboard input and structure-dependent selection}, researchers have explored multimodal interactions, particularly through pen and sketching tools. These approaches demonstrate that drawing or scribbling can convey spatial intent \textcolor{revisioncolor}{through free-form marks}. Yet, existing LLM-integrated sketching systems predominantly treat sketches as a medium for instruction or generation ~\cite{yen2025code, huang2025sketchgpt}---functioning as spatial commands for structural edits or visual output rather than facilitators of open-ended exploration.

\textcolor{revisioncolor}{Compared with these editing- and generation-oriented systems, fewer studies have examined how partial pen input can support inquiry formulation during active reading.} Our work addresses this gap by introducing an interaction technique dedicated to the act of questioning. By transforming free-form ink into contextually meaningful targets, our approach enables learners to designate and refine complex inquiries without the structural rigidity of traditional digital selection.

\section{Formative Study: Understanding Barriers in Standard Workflow}
\label{sec:Formative_Study}

We conducted a formative study to investigate learners' standard workflow for asking questions to LLMs during pen-based learning, identify the fundamental barriers inherent in these workflows, and explore the potential of an in-situ pen-based Q\&A system.

\subsection{Study Design}

\paragraph{Participants}
\textcolor{revisioncolor}{We recruited university students who had used LLMs during pen-based learning within the previous six months ($N=6$; five women and one man; mean age = 24.2 years, range = 23--25; four undergraduate and two graduate students). All participants had used LLMs for at least one year during learning activities such as concept clarification, summarization, question generation, and mathematical or coding assistance. They typically took notes with a stylus while attending classes or studying on a tablet and used an LLM on a laptop when asking questions.}

\vspace{-2mm}
\paragraph{Materials}
To ensure that participants would naturally generate a diverse range of pen-based questions, we selected study materials based on two key criteria: (1) coverage of a wide range of knowledge domains and (2) inclusion of varied content layouts (e.g., figures, tables, graphs, code snippets, and formulas) designed to elicit different pen-based questioning strategies. Based on these criteria, we searched across a broad collection of OpenStax\footnote{\url{https://openstax.org/}} textbooks.
Following this process, we selected four subjects—history, data science, algebra, and biology. During the study, participants read sections from the four subjects sequentially, spending approximately 10 minutes on each. To help them quickly immerse themselves in the content, we also provided a brief supporting document for each subject outlining the learning context (e.g., chapter introduction, summary of preceding content), learning objectives, and a few example questions.

\vspace{-2mm}
\paragraph{Procedure}
We designed the study procedure to observe how learners naturally formulate questions through pen input.
Each study session lasted approximately 60 minutes and was divided into three parts:
\begin{enumerate}
    \item \textbf{Pre-Study Interview (10 min):} Participants discussed their typical pen usage patterns during pen-based learning, the types of systems they use when asking questions to LLMs, and any inconveniences they had experienced.
    \item \textbf{Task Performance (40 min):} Participants engaged in a learning task within a pen-based environment, where they read the provided materials and asked questions. 
    \item \textbf{Post-Study Interview (10 min):} Participants reflected on their overall experience and discussed how it differed from their typical questioning workflow.
\end{enumerate}

\paragraph{Protocol}
During the task, we employed a Wizard-of-Oz~\cite{kelley1984iterative} protocol for the pen-based Q\&A task.
Participants used an iPad and an Apple Pencil to read and annotate PDF documents, with the screen displaying both the document and a web-based chat interface. To enable real-time observation and response simulation, their screen was shared with the researcher via Zoom.
As illustrated in Appendix~\ref{app:formative_study_procedure}, participants asked questions by sketching (e.g., underlines, circles, arrows, or keywords) with a green pen or highlighter and appending a question mark to signal submission. The Wizard captured the sketched region from the live stream, forwarded it to GPT-4o \cite{openai_gpt4o_2024}, and returned the model's response via the chat interface. Participants were able to retrieve and view the answer by refreshing the page. 
A predefined system prompt was used to ensure that the LLM accurately interprets the intentions of the participants from the shared images and pen annotations.

\vspace{-1mm}
\paragraph{Data Collection and Analysis} 
\textcolor{revisioncolor}{We collected screen and audio recordings, interview transcripts, observational notes, and participants’ question sketches.}

\subsection{Findings}

\textcolor{revisioncolor}{Pre-study interviews showed that participants commonly annotated study materials on a tablet but switched to a separate keyboard-equipped device for LLM queries, often transferring context through text descriptions, copy-and-paste, or screenshots. This formative finding later informed the baseline condition in Study 1.}

Through thematic analysis of the interview transcripts and observational data, we identified two barriers that hinder effective question-asking when learners rely on standard workflow where they use a keyboard on a separate device during pen-based learning: \RB~and \EB.

\begin{figure}[!htbp]
    \centering
    \includegraphics[width=\linewidth]{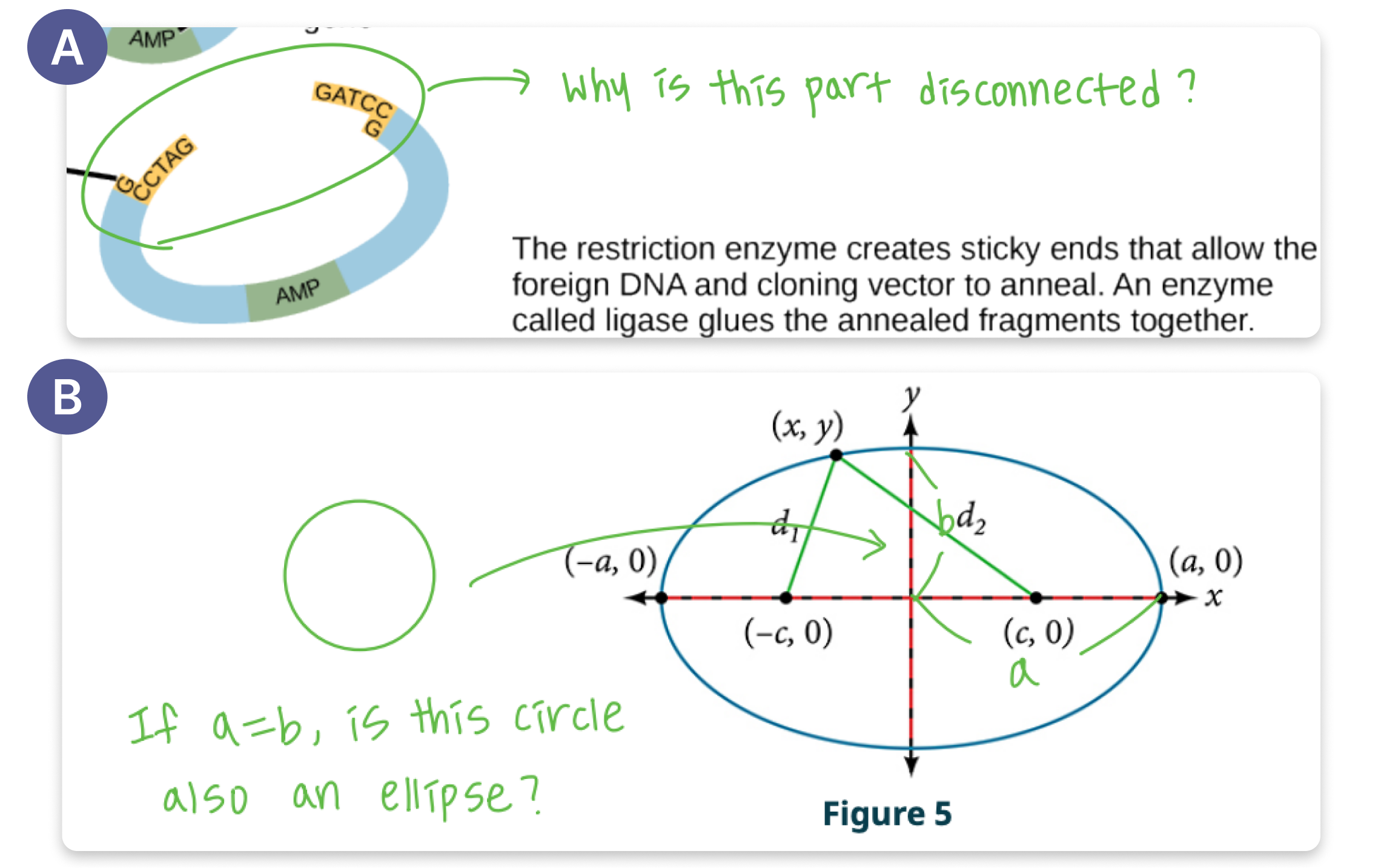}
    \caption{Pen-based questioning strategies observed in the formative study. (A) Using circles to reference specific areas. (B) Drawing sketches to express questions.}
    \label{fig:formative_study}
    \vspace{-2mm}
\end{figure}

\paragraph{Finding 1: \RB---Difficulty in Clearly Referencing Specific Regions}
In standard workflow, participants found it challenging to precisely indicate which part of the study material their question referred to. As a result, they often resorted to inconvenient workarounds, such as repeatedly capturing screenshots or copying and pasting portions of the document. For instance, P2 noted that he typically has to ``\textit{capture specific parts and insert the image or copy and paste}'' whenever he needs to reference particular content within the study material. 
In contrast, with pen-based interaction, participants could naturally sketch to reference specific elements. As shown in Figure~\ref{fig:formative_study}(A), we observed that participants used circles or underlines to reference specific areas. \textcolor{revisioncolor}{They also used familiar annotation practices such as arrows, highlights, and handwritten keywords to link questions to their targets. These observed practices informed \penquiry's interactions rather than defining a fixed one-to-one gesture vocabulary.}

\paragraph{Finding 2: \EB---Limitations of Textual Expression for Visual, Symbolic, and Structural Information}
Participants also noted that text alone was insufficient to fully express visual, symbolic, and structural aspects of their questions. For instance, P5 described the difficulty of inputting mathematical formulas, stating that ``\textit{it is difficult to type formulas with a keyboard when studying mathematics.}'' As a result, he often had to ``\textit{write the formula with a pen on an iPad, capture it, and send the image to GPT}'' whenever he needed to ask a formula-related question.
Conversely, participants using pen input not only naturally wrote formula but also expressed their questions by drawing sketches, as shown in Figure~\ref{fig:formative_study}(B). \textcolor{revisioncolor}{These observations indicate that the barrier involved translating spatial and symbolic intent into linear textual descriptions, rather than merely composing the question itself.}

\vspace{-2mm}
\paragraph{Design Requirements} 
\textcolor{revisioncolor}{These findings led to three design goals: (1) support flexible pen marks for direct reference and visual or symbolic expression, (2) capture the surrounding semantic scope and document context of concise sketches, and (3) preserve spatial and conversational continuity between questions and answers. In particular, participants’ preference for locating responses in the closest available page margin informed the sticky-note-style Answer Shape, while their requests for consecutive questioning motivated the in-place Re-Ask interaction.}
\section{\penquiry~V1: Addressing Barriers of Standard Workflow}
\label{sec:penquiry_v1}

To address \RB~and \EB~identified in our formative study, we designed and implemented \penquiry~V1, an in-situ pen-based Q\&A system. 

\subsection{User Interface and Interactions}

\begin{figure*}[!t]
    \centering
    \includegraphics[width=\linewidth]{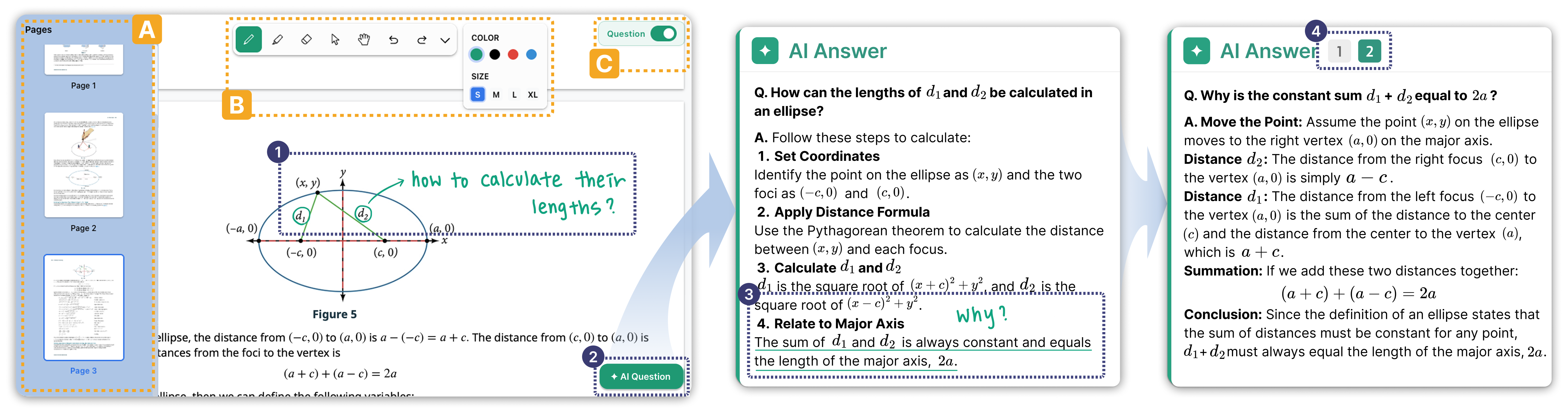}
    \caption{\penquiry~V1 user interface and workflow. The interface features (A) a Thumbnail View, (B) Sketching Tools, and (C) Question Mode. The sequence illustrates the integrated inquiry workflow: (1) sketching an in-situ question, (2) receiving a context-aware in-place answer, (3) sketching a follow-up question on the previous response, and (4) navigating iterative answers via tabs.}
    \label{fig:v1_figure}
\end{figure*}

\subsubsection{Sketch Interface and Interactions}
The \penquiry~V1 interface is designed to support pen-based learning activities directly on top of PDF study materials. It consists of two layers: a Document Layer that renders the original PDF content and a transparent Canvas Layer above it that captures pen input. Using basic sketching tools, learners can freely take notes, highlight passages, and add annotations (Figure~\ref{fig:v1_figure}(B)). They can also navigate the document by performing gestures such as zooming, or by using a thumbnail view (Figure~\ref{fig:v1_figure}(A)). 
The basic sketching tools are implemented using the open-source tldraw library ~\cite{tldraw}.

Beyond these basic interactions, the core feature of \penquiry~V1 is its ability to transform pen-based sketches into queries directed to an LLM. This process is enabled through a dedicated Question Mode, where all sketches are interpreted as question intents (Figure~\ref{fig:v1_figure}(C)).

\subsubsection{Question Creation}
In Question Mode, learners can naturally express their questions by sketching directly on the document. To overcome \RB, learners can precisely indicate their target references using visual markers such as underlines, circles, highlights, and arrows. Furthermore, to address \EB, learners can freely incorporate visual and symbolic expressions---such as mathematical equations, diagrams, or symbols---that are difficult to articulate through text. \textcolor{revisioncolor}{Importantly, sketches created for questioning are explicitly distinguished from regular annotations through this mode; these sketches are temporary and automatically disappear once the Question Mode is deactivated.}

Once one or more question sketches are created, the ``AI Question'' button becomes active (Figure ~\ref{fig:v1_figure}(1, 2)). When tapped, the system generates a query to GPT-4o by combining the sketches with the relevant document content. The LLM's response is displayed as an Answer Shape---a sticky note-like object containing the interpreted question text and the answer. \textcolor{revisioncolor}{To fulfill the design requirement derived from our formative study that the act of inquiry and learning flow must coincide in the same spatial context, the Answer Shape is positioned by default in the document margin closest to the original sketch.} This placement preserves spatial association while avoiding interference with the study material and other elements. Furthermore, learners can manually reposition the Answer Shapes using the selection tool, while a dotted arrow persistently connects the shape to the location where the sketch was created to maintain a clear visual link.

\subsubsection{Re-Ask}
Learners often develop additional questions based on the answers they receive. To support iterative inquiry, \penquiry~V1 enables follow-up questioning. When learners wish to ask further questions related to a previous answer, they can create new sketches directly on top of the Answer Shape (Figure ~\ref{fig:v1_figure}(3)). By then tapping the ``AI Question'' button, the system recognizes this interaction as a follow-up query and automatically incorporates the context of the prior exchange. The LLM's response to the follow-up query is appended as a new page within the existing Answer Shape (Figure ~\ref{fig:v1_figure}(4)). 

\subsection{LLM Input Generation}

\subsubsection{Input Pipeline Implementation}
To ensure the LLM can accurately interpret sketch-based questions, \penquiry~V1 utilizes a multimodal input pipeline. In the formative study, simple visual differentiation of question intents proved problematic due to conflicts with existing content, sketches obscuring underlying text, and ambiguous semantics from insufficient local context. 

To overcome these issues, \penquiry~V1 captures a multimodal input set: a Cropped Annotated Image (containing sketches), a Cropped Original Image (without sketches), and the Context Text explicitly parsed from the PDF document. \textcolor{revisioncolor}{By providing both the image with the sketch and the image without the sketch, the system effectively separates the user's strokes from the surrounding local context.} The two cropped inputs resolve color-conflict and content-obscuration issues, ensuring the user's intent is visible while the original content remains accessible. The Context Text ensures a more precise semantic baseline, addressing ambiguities caused by limited local views. Dedicated system prompts guide the LLM to synthesize these inputs and interpret minimal gestures as meaningful queries. For follow-up actions (Re-Ask), the system provides the relevant cropped region alongside the prior conversation history. The prompts used in the \penquiry~V1 pipeline are provided in Appendix ~\ref{app:penquiry_version1_prompt}.

\subsubsection{Capturing Semantic Scope}
A key challenge for the LLM is inferring the semantic scope of minimal sketches (e.g., a simple line or circle), which often refer to broader content like an entire paragraph or equation. \textcolor{revisioncolor}{To accurately capture this intended scope, the crops generated for the LLM input encompass the direct annotation as well as the local context expanded by a fixed padding area around the sketch.} For general questions, the cropping region is defined relative to the learner's annotation: vertically, it includes the annotation area plus fixed padding; horizontally, it spans the full width of the document. If the annotation extends beyond the document boundaries, the crop automatically expands to encompass the strokes. \textcolor{revisioncolor}{Furthermore, if this cropping region includes a part of a table or figure, it automatically expands to encompass the entire boundaries of that element.} For follow-up actions, the crop defaults to the dimensions of the target Answer Shape and expands if the new annotation exceeds its boundaries.

\section{Study 1: Standard Workflow vs. \penquiry~V1}
\label{sec:study1}

We conducted User Study 1 to compare \penquiry~V1 against standard workflows, guided by three research questions: (RQ1) How effectively does V1 resolve the barriers inherent in standard workflows? (RQ2) What types of questions do learners ask in an in-situ pen environment? and (RQ3) What new, pen-specific barriers emerge when using pen input?

\subsection{Study Design}
\label{sec:study1_study_design}

\begin{figure*}[!t]
    \centering
    \includegraphics[width=\textwidth]{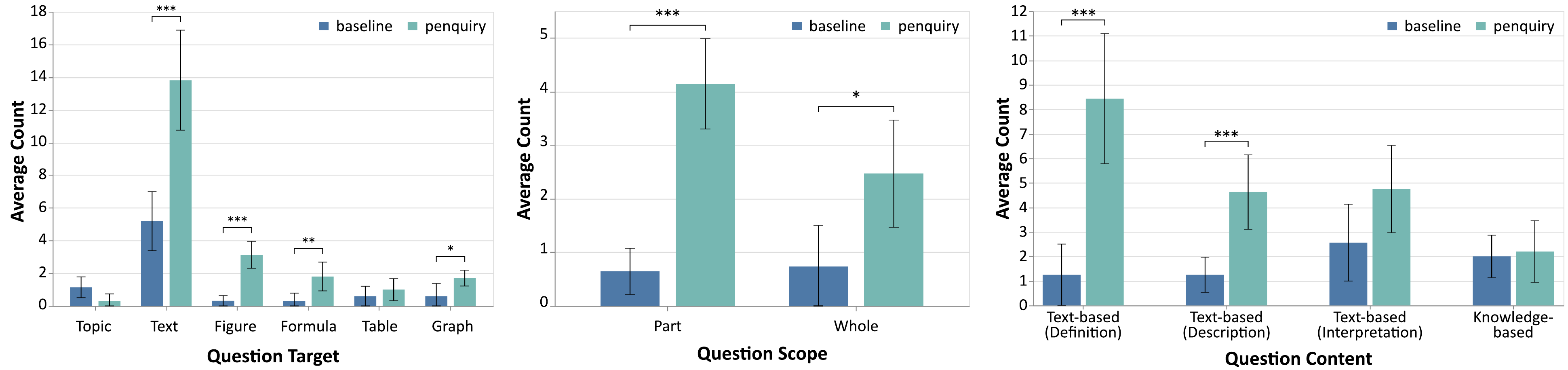}
    \caption{Comprehensive preference analysis across three dimensions. 
    (A) \textbf{Target-based}: \penquiry~V1 was favored for all specific targets, with the exception of general \texttt{Topic} queries. 
    (B) \textbf{Scope-based}: A strong preference for \penquiry~V1 was observed in \texttt{Part}-scoped questions, where spatial referencing is crucial. 
    (C) \textbf{Content-based}: Preference shifted with question complexity; \penquiry~V1 was most effective for short, direct question types such as \texttt{Definition} and \texttt{Description}.}
    \label{fig:result}
\end{figure*}

\paragraph{Participants}
This study was conducted with 16 participants (9 male, 7 female; mean age = 24.8 years). All participants were undergraduate or graduate students with prior experience in both tablet-based pen interactions and using LLMs for academic purposes. To minimize the influence of domain-specific knowledge, we excluded applicants majoring in related fields. Each participant received USD 14 for completing the 75 minute study. All study procedures were approved by our Institutional Review Board (IRB).

\vspace{-2mm}
\paragraph{Materials}
We selected two chapters from a geotechnical engineering textbook in the LibreTexts\footnote{\url{https://eng.libretexts.org/}} library: ``1.5 Standard Penetration Test'' and ``1.7 Cone Penetration Test.'' These chapters were chosen to ensure comparable difficulty, roughly equivalent to a first-year undergraduate elective while minimizing the likelihood of prior familiarity. Both chapters were of similar length and designed to be fully read and understood within a 25 minute session. Each chapter featured a balanced mix of text, figures, tables, graphs, and formulas, ensuring that at least one instance of each element was included to maintain consistency across conditions. 

\vspace{-2mm}
\paragraph{Conditions}
The study was designed to compare two distinct question–answering environments. To ensure ecological validity, we designed the baseline based on our formative study, instantiating the standard workflow where learners study with a pen on a tablet but use a keyboard-equipped device for LLM queries. In the baseline condition, participants read and annotated the document on a tablet, which was implemented by retaining only the basic learning tools from the \penquiry~V1 system. When a question arose, they switched to a separate web browser and typed their query to an LLM (GPT-4o) using a keyboard.
To enable a fair comparison of question-asking capabilities, the baseline also provided access to the same source materials on the LLM. Specifically, participants could copy–paste text from an adjacent browser tab into the LLM input, and were allowed to attach the screenshots as image references. In the \penquiry~V1 condition, participants used the system for all tasks---including reading, note-taking, and asking questions to the LLM---exclusively through pen-based interactions.

\vspace{-2mm}
\paragraph{Apparatus}
The user study was conducted using an Apple iPad Air running iPadOS 18, paired with an Apple Pencil. For the Baseline condition, the keyboard-based environment was provided using a MacBook Air. 

\vspace{-2mm}
\paragraph{Design}
We employed a within-subjects design~\cite{charness2012experimental} in which each participant experienced both experimental conditions. To mitigate potential order effects, conditions were counterbalanced~\cite{brooks2012counterbalancing}.

\vspace{-2mm}
\paragraph{Procedure}
The experiment lasted approximately 75 minutes and followed this sequence:
\begin{enumerate}
    \item \textbf{Briefing and Consent (10 min):} We explained the purpose and procedure and obtained informed consent.
    \item \textbf{First Learning Session (25 min):} Participants studied the first material under their assigned initial condition (Baseline or \penquiry~V1), asking questions as needed.
    \item \textbf{Break (5 min)}
    \item \textbf{Second Learning Session (25 min):} Participants studied the second material under the remaining condition.
    \item \textbf{Post-study Activities (10 min):} Participants completed in-depth interviews and questionnaires evaluating their experiences and the usability of both systems.
\end{enumerate}

Each learning session was limited to 25 minutes, following the principles of the Pomodoro Technique~\cite{gobbo2008pomodoro}, to maintain participants’ concentration and minimize fatigue. All sessions were screen- and audio-recorded with participants' consent for subsequent analysis.

\subsection{Result}

A total of 429 questions were generated throughout the user study, with 225 (52.45\%) were generated in the baseline condition, and 204 (47.55\%) in the \penquiry~V1 condition. Regarding follow-up questions, 49 (21.78\%) were recorded in the baseline condition, compared to 43 (21.08\%) in the \penquiry~V1 condition.

\subsubsection{System Usability}
To compare the user experience across the two conditions, we analyzed participants’ responses using three measures: the System Usability Scale (SUS)~\cite{brooke1996sus}, the Technology Acceptance Model (TAM)~\cite{davis1989perceived}, and the NASA-Task Load Index (NASA-TLX)~\cite{hart1988development} to evaluate the usability of \penquiry~V1. All questionnaire items were rated on a 7-point Likert scale, with statistical significance denoted as *$p < .05$, **$p < .01$, and ***$p < .001$.

\paragraph{SUS}
The average SUS score for the \penquiry~V1 condition was \textbf{80.78}, which corresponds to an ``Excellent'' usability rating according to standard SUS benchmarks.

\vspace{-2mm}
\paragraph{TAM}

\begin{table}[h]
\centering
\caption{Results from the TAM}
\label{tab:tam}
\resizebox{\columnwidth}{!}{
\begin{tabular}{l c c c}
\toprule
\textbf{Question} & \textbf{Baseline} & \textbf{\penquiry~V1} & \textbf{p-value} \\
\midrule
1. This system is useful for learning. & 5.98 & 6.63 & 0.029* \\
2. This system improves learning efficiency. & 5.28 & 6.25 & 0.011* \\
3. The functions of this system are intuitive. & 5.88 & 6.38 & n.s. \\
4. I intend to continue using this system. & 5.44 & 6.31 & 0.008** \\
5. I feel positive about using this system. & 5.69 & 6.50 & 0.007** \\
\bottomrule
\end{tabular}%
}
\end{table}

The evaluation of TAM showed that \penquiry~V1 received significantly higher ratings than the baseline condition in most categories, including perceived usefulness and intention to use (Table \ref{tab:tam}). This suggests that participants rated \penquiry~V1 as a more helpful, efficient, and easier-to-use tool for learning, and also expressed a sustained intention to use the system.

\paragraph{NASA-TLX}

\begin{table}[h]
\centering
\caption{Results from the NASA-TLX}
\label{tab:tlx}
\resizebox{\columnwidth}{!}{
\begin{tabular}{l c c c}
\toprule
\textbf{Question} & \textbf{Baseline} & \textbf{\penquiry~V1} & \textbf{p-value} \\
\midrule
1. Mental Demand & 4.00 & 3.13 & n.s. \\
2. Physical Demand & 4.06 & 2.31 & <0.001*** \\
3. Temporal Demand & 3.81 & 3.69 & n.s. \\
4. Performance & 5.00 & 5.50 & n.s. \\
5. Maintained Learning Flow & 4.06 & 5.63 & 0.046* \\
\bottomrule
\end{tabular}%
}
\end{table}

Compared to the baseline, the results of the NASA-Task Load Index (NASA-TLX) for Cognitive Load and Learning Flow indicated that the \penquiry~V1 condition overall reduced the cognitive burden of performing tasks, particularly by imposing significantly lower physical demand. Through this reduction in cognitive burden, \penquiry~V1 showed a statistically significant advantage in Maintained Learning Flow as well (Table \ref{tab:tlx}).

\subsubsection{Addressing \RB} 

\paragraph{Question Target-based Analysis}
We classified the targets of participants' questions into \texttt{topic}, \texttt{text}, \texttt{figure}, \texttt{formula}, \texttt{table}, and \texttt{graph} (Appendix ~\ref{app:example_study1}(A)). As shown in the preference data (Figure~\ref{fig:result}(A)), participants generally preferred using \penquiry~V1 when asking questions with specific, concrete referents. However, for questions targeting a general \texttt{topic} that lack a clear deictic target, participants favored the baseline condition.

\vspace{-2mm}
\paragraph{Question Scope-based Analysis}
We classified the scope of the targeted elements into \texttt{Whole} (e.g., an entire figure or table) and \texttt{Part} (e.g., a specific axis of a graph or a row in a table) (Appendix ~\ref{app:example_study1}(B)). According to the preference data (Figure~\ref{fig:result}(B)), participants showed an overwhelming preference for using \penquiry~V1 when asking these \texttt{Part}-scoped questions.

\vspace{-2mm}
\paragraph{Pen-Mark Type}
Based on Marshall's annotation classification \cite{10.1145/263690.263806}, we analyzed 204 queries in the \penquiry~V1 dataset. Participants predominantly used four pen-mark types: \texttt{underlining} (45\%), \texttt{circles} (29\%), \texttt{boxes} (9\%), and \texttt{braces} (2\%) (Appendix ~\ref{app:pen_mark}). Notably, 85\% of queries contained at least one visible mark. This overwhelming reliance on visual deictic cues demonstrates that pen input effectively supports precise referencing behaviors.

\vspace{-2mm}
\paragraph{Ease of Referencing and Qualitative Insights}

\begin{figure}
    \centering
    \includegraphics[width=\linewidth]{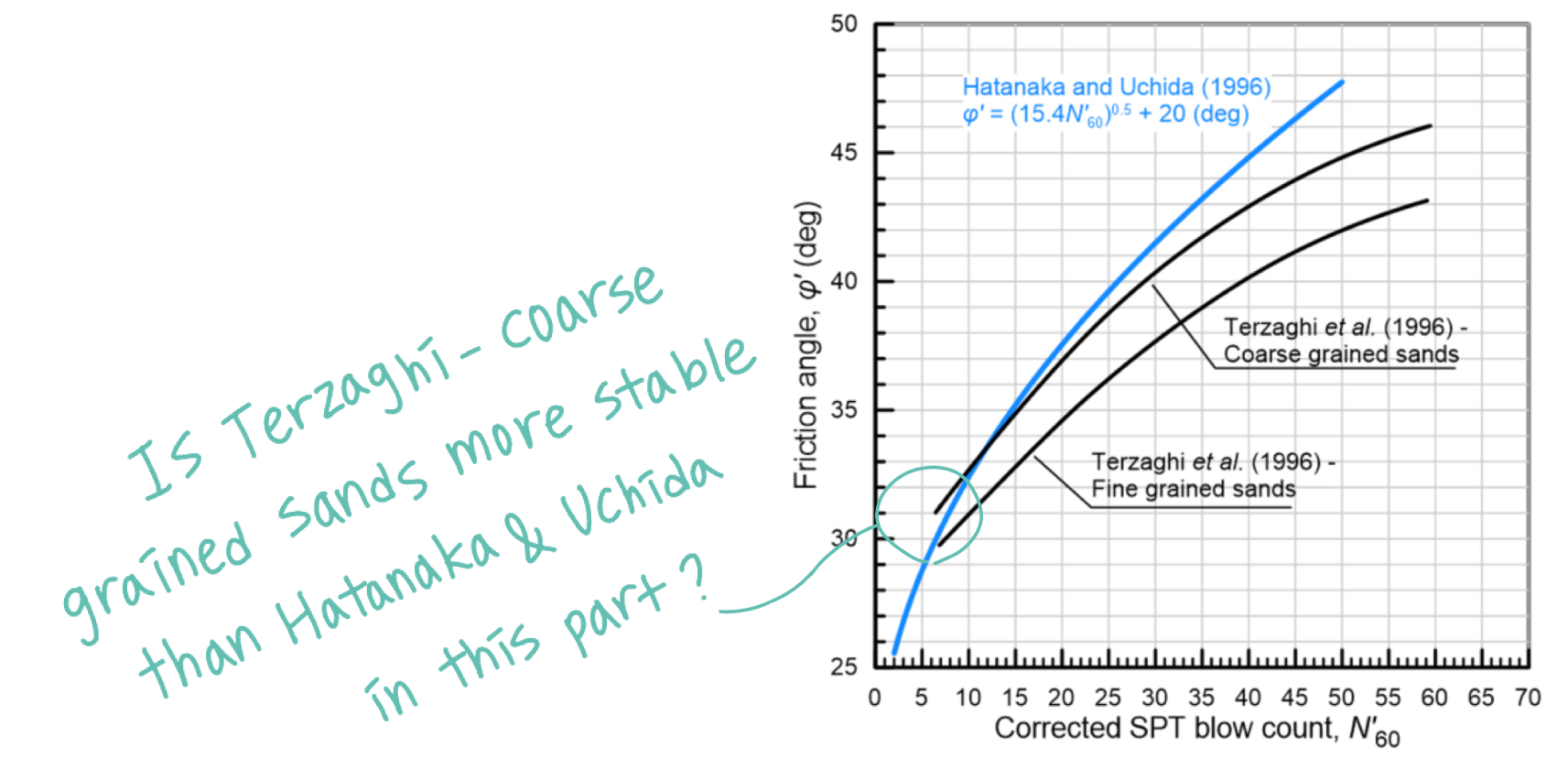}
    \caption{Directly referencing an area difficult to describe textually.}
    \label{fig:result_pen_specific}
\end{figure}

Regarding the ``Ease of referencing specific elements,'' \penquiry~V1 showed a statistically significant improvement over the baseline (6.00 vs. 4.38, p = 0.015*), allowing participants to easily designate specific targets. Interview responses further supported this efficiency. For instance, Figure~\ref{fig:result_pen_specific} demonstrates a case where pointing with a pen is essential. When asking about a specific part of a graph, P16 noted: ``\textit{If I had tried to write this question with text, it would have become much longer, like `between 5 and 10 on the x-axis...' With the pen, I could just point and ask directly.}'' Collectively, these results demonstrate that \penquiry~V1 successfully introduces the principle of deictic reference \cite{bolt1980put} into the Q\&A process. The ability to ``just point and ask directly'' provides unambiguous, in-situ interaction, effectively resolving the referential barrier.

\subsubsection{Addressing \EB}
\label{sec:study1_eb}

\paragraph{Question Content-based Analysis}
We conducted a question content-based analysis to categorize the underlying semantic intent of the inquiries. To assess the cognitive depth of participants' questions, we adopted the classification framework proposed by Scardamalia and Bereiter \cite{scardamalia1992text}. This framework categorizes questions into \textit{Text-based} questions, which seek to understand explicitly presented information, and \textit{Knowledge-based} questions, which involve higher-level cognitive processes such as extending, applying, or comparing beyond the text. We further sub-divided \textit{Text-based} questions into \texttt{Definition} (seeking the meaning of a short term), \texttt{Description} (requesting a simple explanation), and \texttt{Interpretation} (asking about reasons, implications, or principles) (Appendix ~\ref{app:example_study1}(C)). According to the preference data (Figure \ref{fig:result}(C)), participants showed a significant preference for \penquiry~V1 when asking shorter questions, such as \texttt{Text-based (Definition)} and \texttt{Text-based (Description)}. In contrast, as questions became more interpretive or required longer, more elaborate descriptions, a trend toward preferring the baseline emerged. Interviews further clarified this shift; several participants (e.g., P1, P6, P11, P13) reported that typing longer questions was faster and more convenient with a keyboard, leading them to favor the baseline in such cases.

\vspace{-4mm}
\paragraph{Free-form Pen Input}

\begin{figure}[!htbp]
    \centering
    \includegraphics[width=\linewidth]{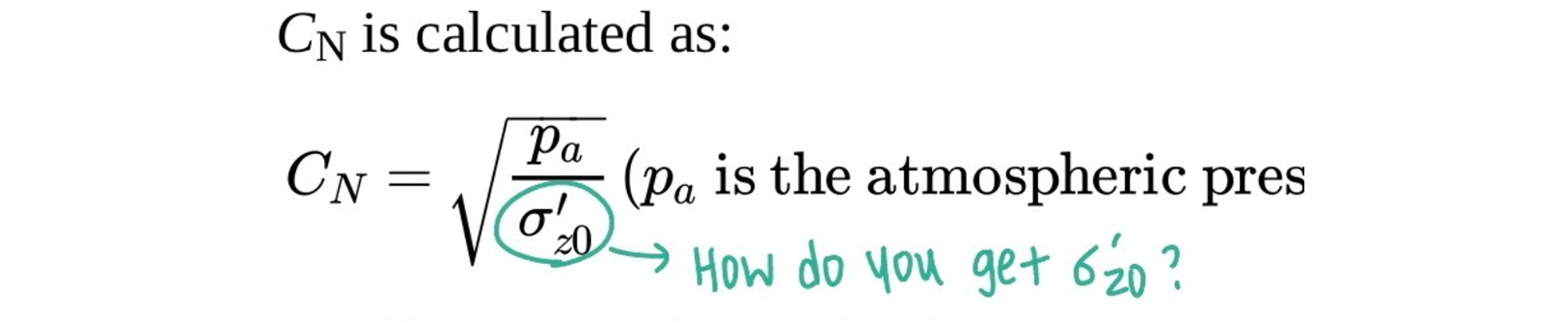}
    \caption{Example of free-form pen-based inquiries. Participants comfortably articulated mathematical symbols and notations using pen input.}
    \label{fig:result_freeform}
\end{figure}

Our analysis examined specific inquiry types where free-form pen input offered a distinct expressive advantage over keyboard modalities. As illustrated in the representative examples in Figure \ref{fig:result_freeform}, participants leveraged the pen to effortlessly articulate mathematical symbols that would otherwise be cumbersome to express via keyboard. These cases confirm that free-form pen input significantly reduces the barrier to entry for visual, structural, or symbolic inquiries, enabling participants to express these complex intents directly.

\vspace{-2mm}
\paragraph{Ease of Expressing Intent and Qualitative Insights}
Regarding the ``Ease of expressing question intent,'' the \penquiry~V1 score was slightly lower than the baseline (4.88 vs. 5.50, p = 0.374), primarily due to difficulties in handwriting long questions or occasional recognition issues. However, this difference was not statistically significant. In fact, interview responses revealed that \penquiry~V1 provided distinct advantages for queries that are difficult to articulate in text. As shown in Figure \ref{fig:result_freeform}, several participants (e.g., P4, P9) expressed a clear preference for the pen modality when handling symbolic and notational content, noting that ``\textit{entering mathematical formulas via a keyboard presents many difficulties, making the pen an easier alternative.}'' They also appreciated the system's accurate recognition of handwritten formulas. Similarly, P10 remarked, ``\textit{With \penquiry~V1, I could easily express my intent by directly denoting elements related to the x-axis or y-axis.}''

\subsection{Finding}
While \penquiry~V1 mitigated the inherent barriers of the standard workflow, our analysis revealed that pen-based interaction introduces its own set of challenges. We redefined \RB~and \EB~within the context of pen-based inquiry as follows.

\paragraph{Finding 1: \RB---Lack of Visual Feedback in Verifying Referential Intent}
Although pen input enables participants to reference materials with high degrees of freedom, this flexibility often introduces uncertainty regarding system recognition. Participants noted a critical lack of visual feedback to confirm whether their intended targets were accurately captured, occasionally resulting in irrelevant system responses. For instance, P14 remarked, ``\textit{I underlined a sentence to ask a question, but the system provided an irrelevant answer. The fact that my pen marks could be misinterpreted made it slightly uncomfortable.}'' Similar frustrations were reported by P1, P15, and P16 when their referential intent was misaligned with the system’s internal processing. These instances underscore that providing explicit visual grounding is essential for users to verify that their intended targets are accurately reflected in the query.

\vspace{-2mm}
\paragraph{Finding 2: \EB---High Interaction Cost in Formulating Complex Queries}
Although pen input successfully supports visually and symbolically rich queries, a significant limitation emerged when participants attempted to formulate \texttt{Knowledge-based} questions. Those who engaged in complex inquiry reported that handwriting lengthy, structured queries was both physically burdensome and cognitively demanding. Furthermore, participants expressed recognition anxiety, fearing that the system might fail to accurately digitize their text even after the effort of writing it out in full. These findings underscore the necessity for an interaction mechanism that can assist learners in articulating lengthy and complex questions, thereby reducing the interaction cost associated with manual handwriting in pen-based environments.

\section{\penquiry~V2: Resolving Pen-Specific Barriers}
\label{sec:penquiry_v2}

Building on the findings from Study 1, we identified residual challenges in pen-based questioning: a lack of immediate confirmation regarding referential accuracy and the significant physical effort required for handwriting extended inquiries. To address these issues and further mitigate \RB~and \EB, we refined the system into \penquiry~V2, which retains the core functionality of V1 while introducing targeted enhancements to the interface interactions and the LLM input generation pipeline.

\subsection{User Interface Enhancements}
\begin{figure}[!htbp]
    \centering
    \includegraphics[width=\linewidth]{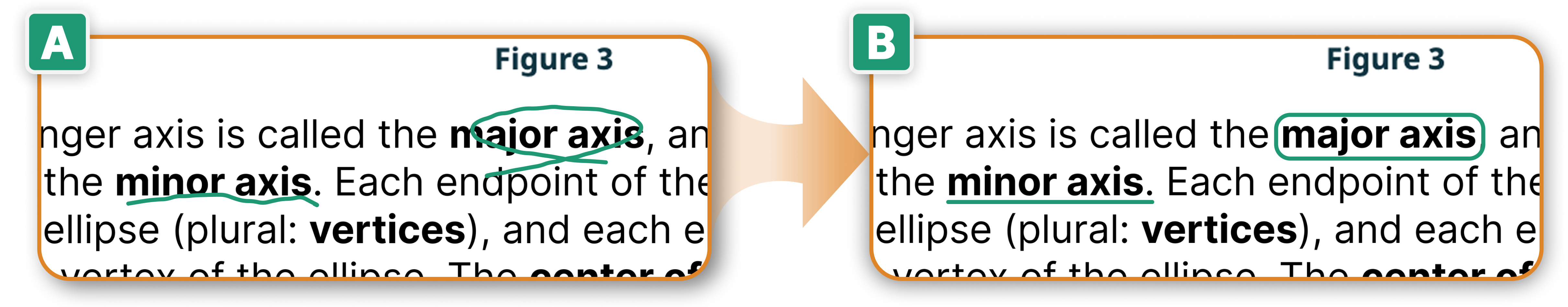}
    \caption{Content Snapping. (A) A learner draws a rough circle around ``major axis'' and underlines ``minor axis''. (B) The system automatically snaps these sketches into precise geometric shapes  (rounded rectangle and line) to enclose the text, providing immediate visual confirmation of the referential target.}
    \label{fig:v2_figure_1}
\end{figure}

\subsubsection{Content Snapping}
To address the uncertainty learners experienced regarding their referenced regions, V2 introduces \contentsnapping. \textcolor{revisioncolor}{When a learner draws a circle or underline over the document (Figure~\ref{fig:v2_figure_1}(A)), the system maps the sketch to nearby character-level PDF text elements and redraws it to enclose or underline those elements (Figure~\ref{fig:v2_figure_1}(B)).} This immediate visual feedback confirms that the system has accurately captured their referential target, eliminating ambiguity before the question is submitted to the LLM. \textcolor{revisioncolor}{This approach to snapping builds on imprecise pen-selection work such as Sloppy Selection \cite{lank2005sloppy}.}

\begin{figure}[!htbp]
    \centering
    \includegraphics[width=\linewidth]{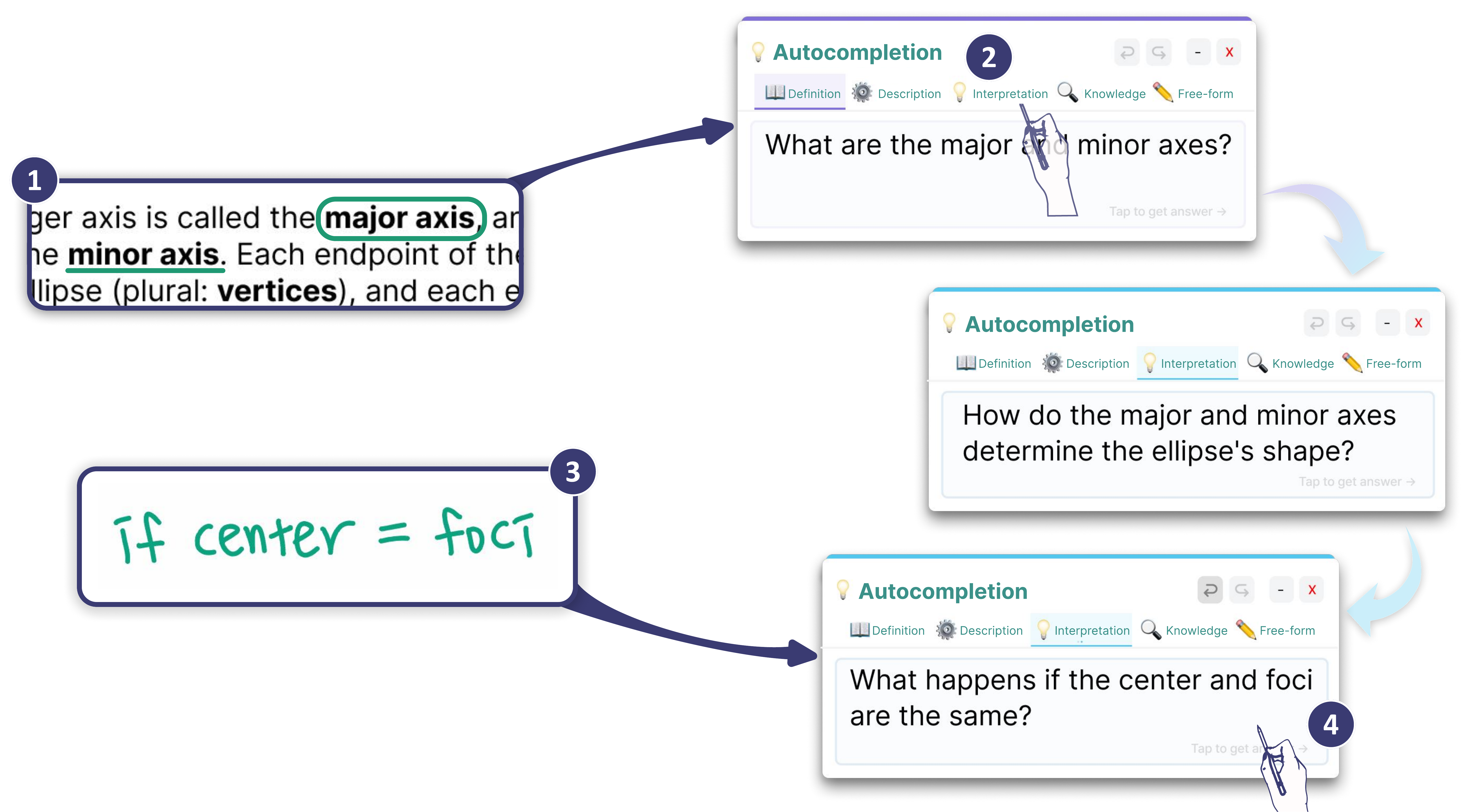}
    \caption{Question Autocompletion. (1) Referencing ``major axis'' and ``minor axis'' generates an autocompleted question under the Definition tab (highest confidence). (2) The learner switches to the Interpretation tab to align with their specific intent. (3) Adding a supplementary sketch dynamically refines the suggested question to incorporate the new context. (4) Tapping the updated question submits the request.}
    \label{fig:v2_figure_2}
\end{figure}

\subsubsection{Question Autocompletion}
To alleviate the physical and cognitive burden of handwriting entire questions, V2 features \questionautocompletion. When a learner provides a minimal sketch---such as a brief underline or a circled term, occasionally accompanied by a keyword---the system analyzes the action within its local context to dynamically generate four autocomplete question candidates (Figure~\ref{fig:v2_figure_2}(1)).
These candidates are categorized based on the question-content analysis detailed in Section~\ref{sec:study1_eb}.

The LLM evaluates the user's sketch to calculate a confidence score for each category, displaying the autocompleted question with the highest confidence. Learners can navigate through alternative categories to find a question that aligns with their specific intent (Figure~\ref{fig:v2_figure_2}(2)). If the initially autocompleted questions do not fully capture their meaning, learners can continue adding sketches to refine them (Figure~\ref{fig:v2_figure_2}(3)), or revert their actions using the undo button. Once satisfied, the learner can tap the autocompleted question to submit it (Figure~\ref{fig:v2_figure_2}(4)).

Furthermore, to preserve user autonomy, V2 maintains the \textit{Free Question} feature from V1. If no autocompleted questions align with their intent, learners can bypass the autocompletion entirely and formulate inquiries through free-form sketching.

\subsection{LLM Input Generation}

\subsubsection{Question Autocompletion and Refinement Pipeline}
To support \questionautocompletion, V2 employs a specialized multi-stage pipeline \textcolor{revisioncolor}{using GPT-4o for all stages}. For the initial \questionautocompletion, the system transmits a multimodal input set to the LLM: the Context Text, a Cropped Original Image, and a Cropped Annotated Image containing the user’s minimal sketch. Using a dedicated prompt, the LLM processes these spatial and semantic inputs to generate four candidate questions, each accompanied by a confidence score.

If a learner refines a suggestion by adding supplementary sketches or keywords, a refinement request is triggered. In this stage, the LLM receives the Selected Category, the Seed Question, the Context Text, and the updated Cropped Original Image and Cropped Annotated Image. The LLM synthesizes these continuous strokes with the existing semantic intent to produce a refined, contextually accurate question. The prompts used across the \penquiry~V2 pipeline are provided in Appendix ~\ref{app:penquiry_version2_prompt}.

\subsubsection{Question Selection and Answering Pipeline}
Once the inquiry is finalized the system proceeds to the final answering stage. The input set for this phase includes the Selected Question, the Context Text, and the corresponding Cropped Original/Annotated Images. The LLM generates a precise, context-aware response, maintaining strict grounding within the surrounding textual and visual boundaries of the document.

\section{Study 2: \penquiry~V1 vs. \penquiry~V2}
\label{sec:study2}

We conducted Study 2 to evaluate the iterative improvements of \penquiry~V2 relative to V1, focusing on two research questions: (RQ1) How effectively does V2 resolve the pen-specific barriers identified in Study 1? and (RQ2) How do learners practically utilize the newly added features during their study?

\subsection{Study Design}
User Study 2 followed the same design and procedure as User Study 1 (Section \ref{sec:study1_study_design}). We recruited a new cohort of 16 participants (7 male, 9 female; mean age = 22.69 years) and employed a within-subjects design to compare two system iterations, \penquiry~V1 and V2. For the study materials, we selected two new chapters (``4. Theory'' and ``5. Carbon'') from a climate science textbook\footnote{\url{https://open.oregonstate.education/}}.

\subsection{Result}
A total of 458 questions were generated across both conditions: 219 (47.8\%) in \penquiry~V1 and 239 (52.2\%) in \penquiry~V2. Follow-up questions accounted for 51 (23.3\%) in V1 and 41 (17.2\%) in V2. 

\subsubsection{System Usability}
We evaluated the usability of \penquiry~V2 using the same three instruments: SUS, TAM, and NASA-TLX.

\vspace{-2mm}
\paragraph{SUS}
\penquiry~V2 achieved an average SUS score of \textbf{84.9}, reflecting an ``Excellent'' usability rating and surpassing the score from the initial study. 

\vspace{-2mm}
\paragraph{TAM}

Both V1 and V2 received consistently high scores across all metrics. While \penquiry~V2 scored slightly higher on items such as perceived usefulness (5.69 vs. 6.19), learning efficiency (5.38 vs. 5.81), intuitiveness (6.25 vs. 6.38), intention to use (5.06 vs. 5.50), and overall positive affect (5.50 vs. 6.06), these differences were not statistically significant. These results indicate that participants found both pen-based systems to be highly intuitive, engaging, and valuable for their learning process.

\vspace{-2mm}
\paragraph{NASA-TLX}
Both systems minimized user burden while maintaining high learning quality. Participants reported low levels of mental (3.19 vs. 3.13), physical (3.63 vs. 2.75), and temporal demand (2.69 vs. 1.94) for V1 and V2, respectively. Conversely, positive metrics such as perceived performance (5.63 vs. 5.75) and learning flow (5.38 vs. 5.25) remained consistently high. Although \penquiry~V2 showed a noticeable trend toward lower physical and temporal demands, the differences between the two versions did not reach statistical significance.

\subsubsection{System Performance and Robustness}

\textcolor{revisioncolor}{
\paragraph{System Error Rate.}
The error rates were 4.6\% (10/219) in \penquiry~V1 and 3.8\% (9/239) in \penquiry~V2. Among the 19 total errors, eight involved handwriting misrecognition and nine involved intent misinterpretation. The remaining two were caused by referential misalignment in V1 and context over-restriction in V2, with the latter resulting from constraining answers strictly to the provided document without external search.}

\vspace{-6mm}
\textcolor{revisioncolor}{
\paragraph{System Latency.}
The average end-to-end response time was 9.88 seconds for the baseline, 8.17 seconds for \penquiry~V1, and 9.04 seconds for \penquiry~V2. For V2, the latency comprised 4.34 seconds for Question Autocompletion and 4.70 seconds for answer generation, remaining comparable to the baseline despite the additional processing stage.}

\subsubsection{Addressing \RB}
\paragraph{Content Snapping}
Participants reported that \contentsnapping~ \newline provided immediate visual confirmation of their intended referents, bridging the gap between action and system perception. While \penquiry~V1 users often felt uncertainty---noting it was ``\textit{not clear if the system accurately captured my range}'' (P5)---the instant alignment in V2 explicitly visualizes the recognition state. From an HCI perspective, this immediate feedback mitigates the Gulf of Execution \cite{norman1986cognitive, park2026bridging, nielsen1994enhancing} by confirming that pointing actions are successfully translated into captured intent. As P4 noted, ``\textit{I could clearly see that what I wanted to ask was being well recognized},'' which allowed participants to formulate questions ``\textit{with trust}'' (P6) and resolved the anxiety (P16) previously caused by ambiguous system feedback.

\vspace{-2mm}
\paragraph{Qualitative Insights on Ease of Referencing}

Both versions received high scores for ``Ease of referencing specific elements'' (5.19 vs. 5.75, $p=0.112$), confirming that pen-based systems are inherently effective for target designation. Beyond simple bounding, the free-form nature of pen input enables users to specify abstract spatial concepts intuitively.
For example, P5 inquired about the area under a curve by simply drawing hatching marks: ``\textit{I wanted to ask about the area under the graph, so I drew hatching to indicate the space, and V2 recommended exactly what I wanted.}'' This illustrates how \penquiry~V2 leverages flexible input to convey complex spatial references that are otherwise cumbersome to describe textually.

\subsubsection{Addressing \EB}

\paragraph{Question Autocompletion}
The effectiveness of the \questionautocompletion~ is evidenced by its high adoption rate, with \textbf{74.9\%} of all submitted queries utilizing autocompletion. While V1 users often found handwriting entire questions physically tiring and time-consuming, V2 resolved this by enabling learners to generate complete inquiries from minimal sketches or keywords. P3 noted that the ability to select autocompleted questions meant they ``\textit{didn't have to write the sentence to the end, which reduced the burden.}'' Furthermore, participants reported that the suggestions reflected their intentions across predefined categories (P5, P12). \textcolor{revisioncolor}{At the same time, retaining the Free Question option allowed them to bypass suggestions when they did not match their intent.}

\begin{figure}[t]
    \centering
    \includegraphics[width=\linewidth]{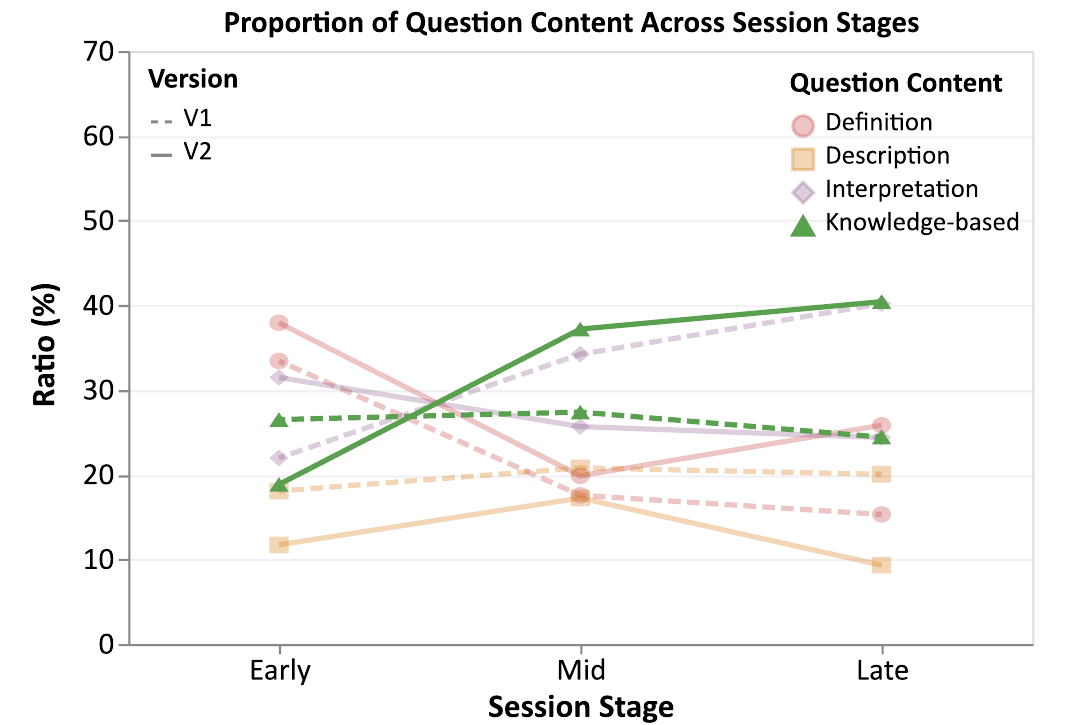}
    \caption{Temporal trends in average proportions of question contents across session phases by system version.}
    \label{fig:result2_kb}
    \vspace{-5mm}
\end{figure}

Beyond reducing physical effort, we investigated whether \questionautocompletion~ enabled learners to formulate more complex, high-level inquiries. Since knowledge-based questions typically emerge in later stages as learners gain deeper comprehension, we analyzed the questioning sessions across three chronological phases: early, mid, and late. Results revealed a significant progressive increase in the proportion of knowledge-based questions: $18.85\%$ (early), $37.19\%$ (mid), and $40.43\%$ (late) (Figure~\ref{fig:result2_kb}; Friedman test, $p = 0.015$; Wilcoxon signed-rank test for early vs. late, $p = 0.019$). 

These findings suggests that the \questionautocompletion~ \textcolor{revisioncolor}{may have supported} an effective cognitive scaffold~\cite{wood1976role}; rather than merely completing sentences, it progressively guided learners toward higher levels of exploration as the session advanced. Interview data corroborated this effect; participants noted that the system prompted inquiries they had not initially considered. \textcolor{revisioncolor}{We also observed cases in which participants first asked their intended question and later returned to a suggestion-inspired follow-up.} 
Ultimately, the feature scaffolded deeper engagement (P1, P3) by lowering the threshold for complex inquiry without imposing additional cognitive load.

\vspace{-2mm}
\paragraph{Qualitative Insights on Ease of Expressing Intent}
Regarding the ``Ease of expressing question intent,'' \penquiry~V2 scored slightly higher than V1 (5.88 vs. 5.50), indicating that both pen-based systems effectively support inquiry conveyance. However, qualitative feedback revealed that V2's \questionautocompletion~ specifically mitigated the cognitive burden of question formulation---a well-documented challenge for learners \cite{king1994guiding}. Participants emphasized that generating full question from minimal keywords significantly lowered this barrier. As P5 noted, the system translated ``\textit{abstract thoughts into well-formed questions, reducing the cognitive burden.}'' P2 echoed this, explaining that even when they held a vague curiosity without a clear explanation, simply writing a keyword like ``VS'' allowed the system to generate an appropriate structured question. Ultimately, this keyword-driven approach alleviated both ``\textit{the burden of making question and the burden of handwriting}'' (P8).

\section{Discussion}
\label{sec:discussion}

\subsection{Unique Affordances of Pen Input Modality}
In this section, we examine the unique affordances of pen-based interaction, focusing on its advantages in referential precision and direct contextual articulation. \textcolor{revisioncolor}{Unlike direct manipulation approaches that operate on predefined document structures, such as text spans or SVG elements \cite{masson2024directgpt}, free-form ink allows learners to indicate structurally undefined yet semantically meaningful regions through deictic marks and free trajectories (Figure~\ref{fig:result_pen_specific} and Figure \ref{fig:result_freeform}).} This in-situ referential expression captures intent seamlessly, without forcing users to translate visual attention into cumbersome textual descriptions. 
\textcolor{revisioncolor}{Furthermore, pen input also supports combining references, symbols, equations, and relational marks within the same canvas, allowing learners to express their inquiry intent in forms closely aligned with the source material. These affordances preserve critical spatial relationships among user marks, document content, and generated responses.}

\subsection{Material-Dependent Adaptive Scaffolding}
While the \questionautocompletion~ feature mitigated the \EB, qualitative feedback revealed that its utility \textcolor{revisioncolor}{scaled dynamically with the learner's domain expertise.} Participants relied heavily on the LLM's generative scaffolding when encountering novel concepts; P12 found the feature ideal when ``\textit{studying a topic for the first time}'' while P9 explained that formulating questions from scratch is especially challenging with unfamiliar material. 
Conversely, the proactive nature of the autocompletion could sometimes hinder learners, with P4 reporting that the automated suggestions occasionally ``\textit{interfered}'' when they already had a very specific question in mind for familiar content. \textcolor{revisioncolor}{Furthermore, generated candidate questions may anchor learners to system-framed perspectives or narrow the range of alternatives they consider, highlighting a potential risk of over-steering.} These insights emphasize that scaffolding should not be a static intervention. 
Consequently, future pen-based Q\&A interfaces must evolve beyond static features to implement adaptive fading \cite{arnold2020predictive, jakesch2023co, bansal2021does}. By intelligently shifting the locus of control---providing aggressive query synthesis during initial conceptual shifts, then fading to subtle, user-driven inquiry as proficiency increases---systems can prevent AI over-reliance while preserving learner agency.

\subsection{Limitation and Future Work}
Despite the promising results, this study has several limitations. First, the limited session duration may have prevented participants from fully adapting to the pen-based questioning environment. The controlled experimental design---requiring the study of unfamiliar materials within a strict timeframe---may also have constrained the diversity of query types. Long-term learning outcomes remain to be explored, highlighting the need for extended longitudinal studies to assess the sustained impact of \penquiry~on learning habits. 

Second, while the baseline in Study 1 was chosen for ecological validity based on prevailing workflows at the time, the rapid evolution of LLM-integrated tools continues to shift standard practices. Future research should incorporate contemporary benchmarks to evaluate \penquiry's distinct advantages within increasingly AI-saturated learning environments.
 
Furthermore, building on our findings regarding cognitive scaffolding, future iterations should explore temporally adaptive \questionautocompletion~ mechanisms. As learners’ comprehension deepens over a session, the system could dynamically adjust the cognitive complexity of its autocompletions based on elapsed time or interaction history. By progressively transitioning from foundational fact-checking in early phases to high-level, knowledge-based prompts in later stages, \penquiry~could provide a truly tailored, adaptive scaffolding experience.
\section{Conclusion}
\label{sec:conclusion}

We identified the \RB~and \EB~that learners encounter when integrating LLMs into pen-based digital learning. To bridge this interaction gap, we presented \penquiry, an in-situ Q\&A system to facilitate inquiry directly within active study workflows. Through an iterative design process, we introduced \contentsnapping~ to resolve referential ambiguity and \questionautocompletion~ to alleviate the physical and cognitive overhead of handwriting queries. Our evaluations demonstrated that these features not only mitigate the limitations of standard keyboard-centric interfaces but also function as effective cognitive scaffolds, guiding learners toward deeper, knowledge-based exploration. Ultimately, \penquiry~establishes a new blueprint for intuitive and effective human--AI interaction, empowering learners to harness the full potential of LLMs without disrupting the cognitive flow of pen-based learning.

\begin{acks}
    This work was supported by the National Research Foundation of Korea (NRF) grants funded by the Korean government (MSIT) (No. NRF-2023R1A2C2005209; No. RS-2024-00456247; No. RS-2023-00218913; No. RS-2025-00520170), the Institute of Information \& Communications Technology Planning \& Evaluation (IITP) grants funded by the Korean government (MSIT) [No. RS-2021-II211343, Artificial Intelligence Graduate School Program (Seoul National University); No. RS-2019-II191906, Artificial Intelligence Graduate School Program (POSTECH); No. RS-2026-25546560, the Leading Generative AI Human Resources Development], and the Hankuk University of Foreign Studies Research Fund. The ICT at Seoul National University provided research facilities for this study.
\end{acks}

\bibliographystyle{ACM-Reference-Format}
\bibliography{ref}

\onecolumn
\appendix

\section{Appendix}

\subsection{Wizard-of-Oz Procedure}
\label{app:formative_study_procedure}
\begin{figure}[H]
    \centering
    \includegraphics[width=\linewidth]{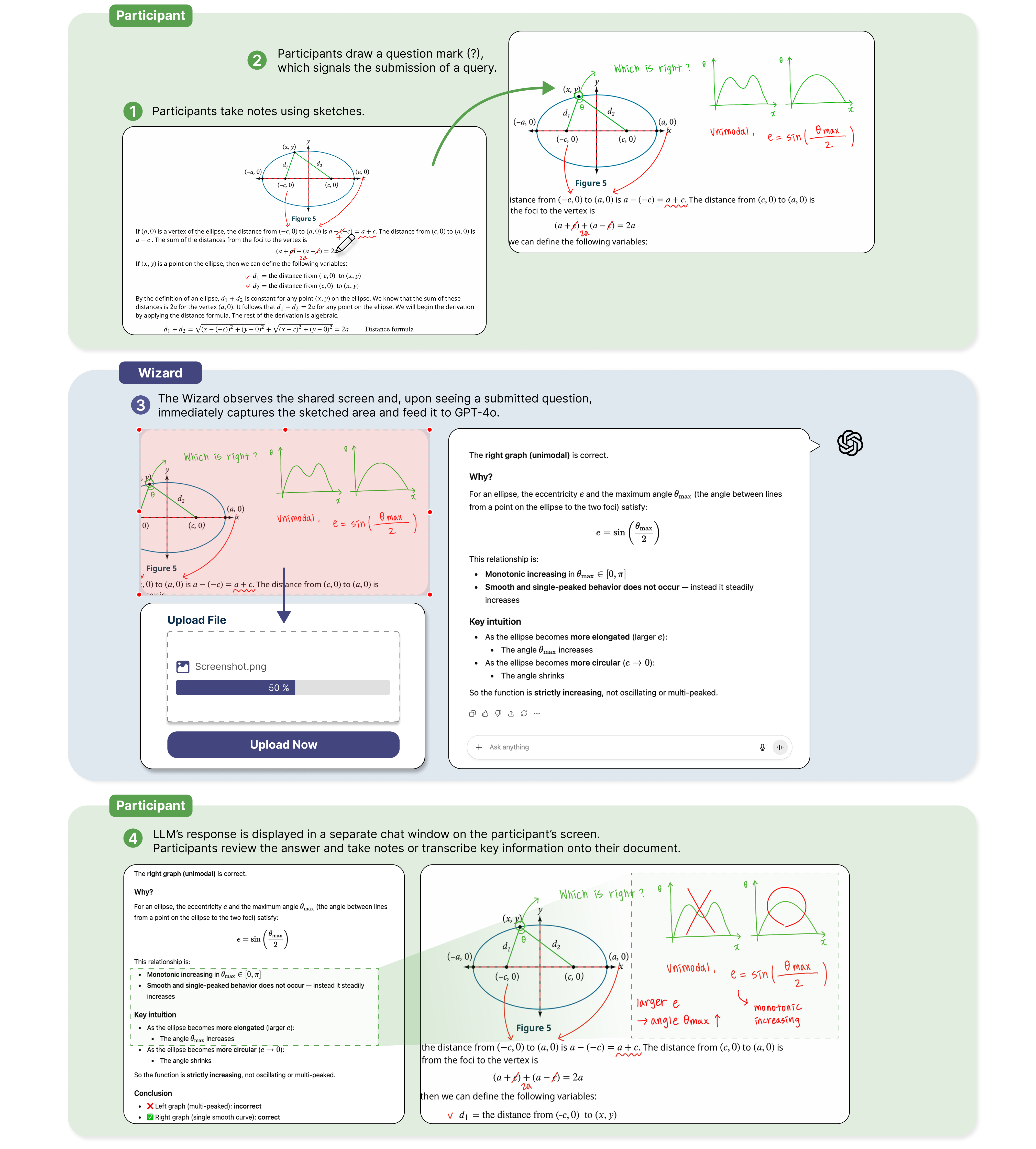}
    \caption{Participants used a green pen or highlighter to sketch and mark their questions. These inputs were captured by the Wizard and forwarded to GPT-4o, whose responses appeared in a separate chat window. Other pen colors were reserved for regular note-taking.}
    \label{fig:wizard_of_oz}
\end{figure}

\subsection{Examples of Question Types}
\label{app:example_study1}
\begin{figure}[H]
    \centering
    \includegraphics[width=\linewidth]{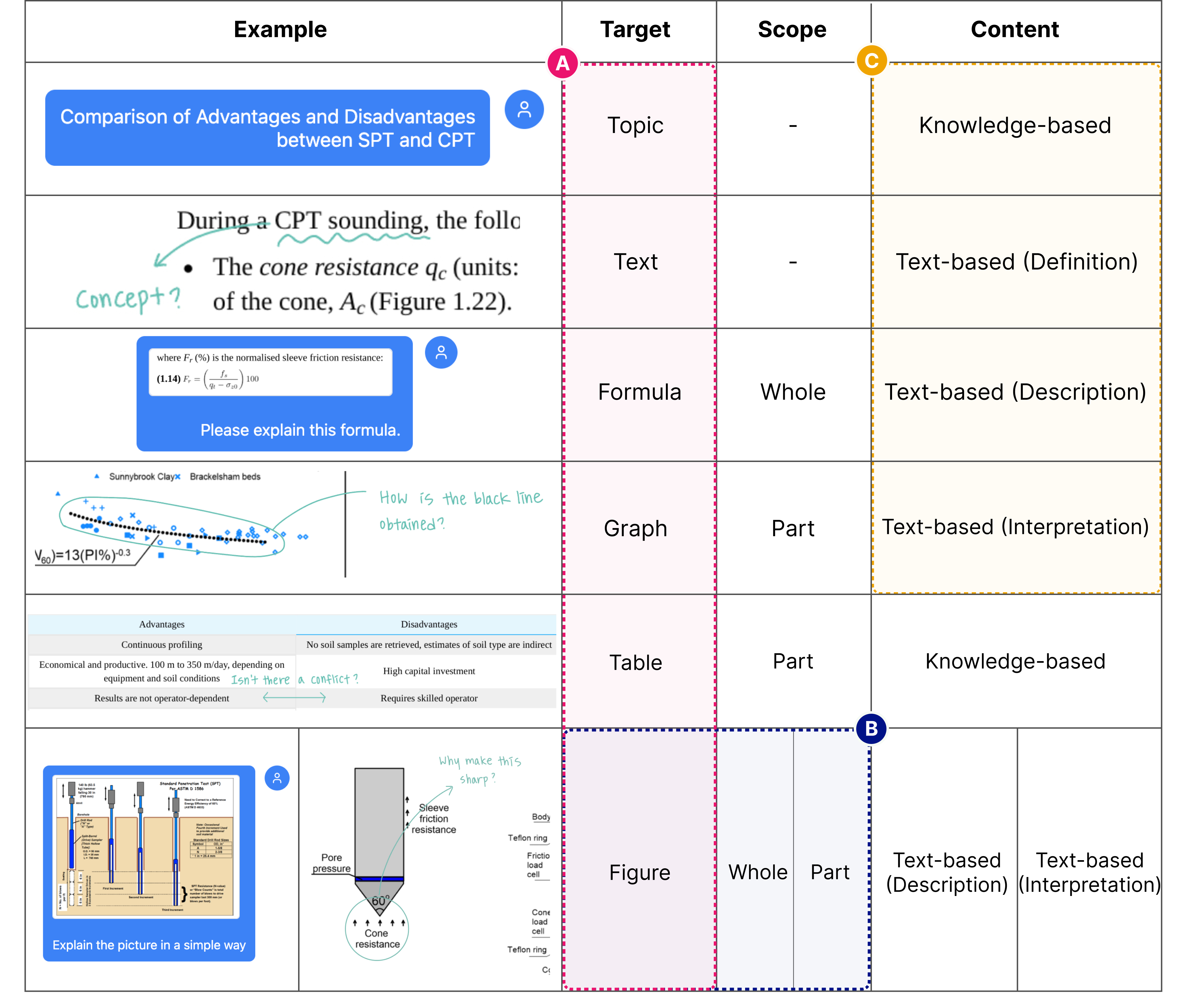}
    \caption{Examples of question Types. (A) Target-based categories: topic, text, formula, graph, table, and figure, (B) Scope-based categories: whole and part, (C) Content-based categories: text-based (definition), text-based (description), text-based (interpretation), and knowledge-based.}
\end{figure}

\subsection{Examples of Pen Mark}
\label{app:pen_mark}
\begin{figure}[H]
    \centering
    \includegraphics[width=\linewidth]{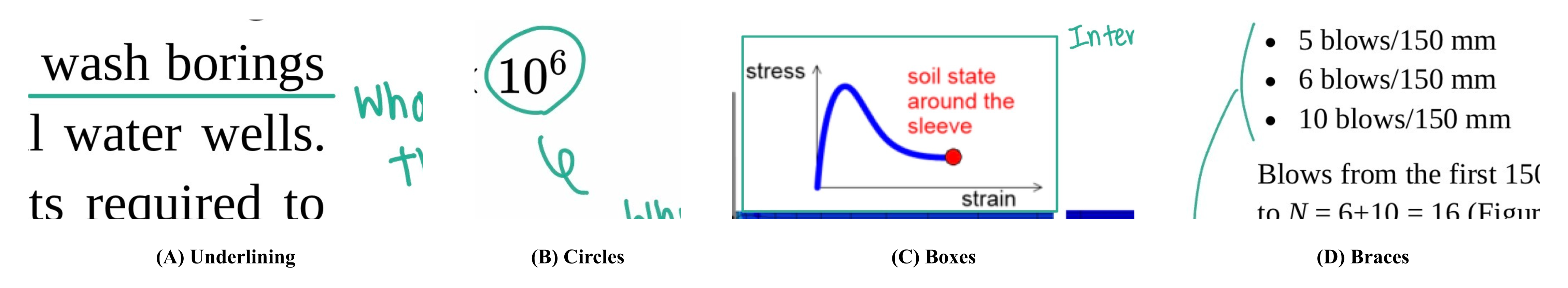}
    \caption{Examples of queries by pen-mark type: (A) Underlining, (B) Circles, (C) Boxes, (D) Braces.}
\end{figure}

\subsection{Formative Study Prompt}
\label{app:formative_study_prompt}

\begin{framed}
 You are an assistant who interprets only the content written in green pen or highlighter as questions in the provided PDF. Any other annotations are not questions. 
For each incoming query, you will be shown an image containing only the region marked in green.
Based on the green-highlighted annotation, reconstruct the intended question and provide an accurate answer.
\end{framed}

\subsection{\penquiry~V1 Prompts}
\label{app:penquiry_version1_prompt}

\subsubsection{AI Question Prompt}
\label{app:ai_question_prompt}
\leavevmode\par\nobreak

\begin{framed}
\begingroup
\small

\noindent\textbf{[ Role ]} \\
You are an assistant who accurately reconstructs the question intent through pen-annotation comparison and provides precise and concise answers based on the context of the page.
\begin{itemize}[leftmargin=1.2em, labelsep=0.4em, itemsep=0pt, parsep=0pt]
    \item Use the provided \texttt{context\_text} to understand the context of the page.
    \item Align the two images and compare them to identify newly added sketches/handwriting (the question), then link them precisely to the surrounding text/figures/equations.
\end{itemize}

\vspace{0.5em}
\noindent\textbf{[ Inputs ]}
\begin{itemize}[leftmargin=1.2em, labelsep=0.4em, itemsep=0pt, parsep=0pt]
    \item \texttt{context\_text}: Original text of the page where the sketch is located.
    \item \texttt{region\_image\_before}: Original image of the region before the question.
    \item \texttt{region\_image\_after}: Image of the region after the question (including sketches).
\end{itemize}

\vspace{0.5em}
\noindent\textbf{[ Core Rules ]}
\begin{itemize}[leftmargin=1.2em, labelsep=0.4em, itemsep=0.2em, parsep=0pt]
    \item[(1)] \textbf{Registration \& Difference Extraction:} Align \texttt{image\_before} $\leftrightarrow$ \texttt{image\_after} and extract $\Delta$sketches (lines, arrows, circles, underlines, symbols, characters).
    \item[(2)] \textbf{Interpretation of Sketch Meaning:}
    \item[] \hspace{0.8em}\textbullet\ Arrow/pointing $\rightarrow$ extract the indicated target.
    \item[] \hspace{0.8em}\textbullet\ Circle/underline $\rightarrow$ treat as indicated/emphasized target and recognize only that range as the question area.
    \item[] \hspace{0.8em}\textbullet\ Question mark / ``why/how/?'' $\rightarrow$ classify question type.
    \item[] \hspace{0.8em}\textbullet\ Short words/equations $\rightarrow$ reinforce with keywords/symbols/variables.
    \item[] \hspace{0.8em}\textbullet\ Sketches may be in Korean, English, or symbols.
    \item[(3)] \textbf{Context Connection:} Find the OCR block/caption/equation that the sketch points to or includes, and refine meaning using the whole page context. Use external knowledge only as a supplement.
    \item[(4)] \textbf{Question Reconstruction:} Summarize the question in one sentence reflecting sketch position/expression. It must end with a question mark.
    \item[(5)] \textbf{Answer Construction:} Provide an accurate explanation based on the document context. Add simple examples/intuitions/steps if needed.
    \item[(6)] \textbf{Formula/Figure Notation:} Write formulas in LaTeX where possible. Refer to diagrams/tables explicitly (e.g., ``region (b) of Figure n'').
    \item[(7)] \textbf{Safety/Quality:} If not present in the textbook context, use external knowledge. Do not expose chain of thought (only provide conclusions and key grounds).
\end{itemize}

\vspace{0.5em}
\noindent\textbf{[ Outputs ]}
\par\noindent\textbf{Q.} [One-sentence question summary reflecting sketch position and intent. Must end with '?' and be bolded.] \\
\textbf{A.} [A complete answer consistent with the document context (with examples, formulas, or references if needed).]

\vspace{0.5em}
\noindent\textbf{[ Style ]}
\begin{itemize}[leftmargin=1.2em, labelsep=0.4em, itemsep=0pt, parsep=0pt]
    \item Use LaTeX for formulas where possible.
\end{itemize}

\endgroup

\end{framed}

\subsubsection{Re-Ask Prompt}
\label{app:re-ask_prompt}
\leavevmode\par\nobreak

\begin{framed}
\begingroup
\small

\noindent\textbf{[ Role ]} \\
You are an assistant who accurately reconstructs the question intent through pen-annotation comparison and provides precise and concise answers based on the given answer text.

\vspace{0.5em}
\noindent\textbf{[ Inputs ]}
\begin{itemize}[leftmargin=1.2em, labelsep=0.4em, itemsep=0pt, parsep=0pt]
    \item \texttt{region\_image\_before}: Original image of the region before the additional question.
    \item \texttt{region\_image\_after}: Image of the region after the additional question (including sketches).
    \item \texttt{original\_answer\_text}: Original LLM answer text referenced by the additional question.
\end{itemize}

\vspace{0.5em}
\noindent\textbf{[ Core Rules ]}
\begin{itemize}[leftmargin=1.2em, labelsep=0.4em, itemsep=0.2em, parsep=0pt]
    \item[(1)] \textbf{Registration \& Difference Extraction:} Align \texttt{image\_before} $\leftrightarrow$ \texttt{image\_after} and extract $\Delta$sketches (lines, arrows, circles, underlines, symbols, characters).
    \item[(2)] \textbf{Interpretation of Sketch Meaning:}
    \item[] \hspace{0.8em}\textbullet\ Arrow/pointing $\rightarrow$ extract the indicated target.
    \item[] \hspace{0.8em}\textbullet\ Circle/underline $\rightarrow$ treat as indicated/emphasized target and recognize only that range as the question area.
    \item[] \hspace{0.8em}\textbullet\ Question mark / ``why/how/?'' $\rightarrow$ classify question type.
    \item[] \hspace{0.8em}\textbullet\ Short words/equations $\rightarrow$ reinforce with keywords/symbols/variables.
    \item[] \hspace{0.8em}\textbullet\ Sketches may be in Korean, English, or symbols.
    \item[(3)] \textbf{Context Connection:} Find the OCR block/caption/equation that the sketch points to or includes, and refine meaning using the whole page context. Use external knowledge only as a supplement.
    \item[(4)] \textbf{Question Reconstruction:} Summarize the question in one sentence reflecting sketch position/expression. It must end with a question mark.
    \item[(5)] \textbf{Answer Construction:} Provide an accurate explanation based on the document context. Add simple examples/intuitions/steps if needed.
    \item[(6)] \textbf{Formula/Figure Notation:} Write formulas in LaTeX where possible. Refer to diagrams/tables explicitly (e.g., ``region (b) of Figure n'').
    \item[(7)] \textbf{Safety/Quality:} If not present in the textbook context, use external knowledge. Do not expose chain of thought (only provide conclusions and key grounds).
\end{itemize}

\vspace{0.5em}
\noindent\textbf{[ Outputs ]}
\par\noindent\textbf{Q.} [One-sentence question summary reflecting sketch position and intent. Must end with '?' and be bolded.] \\
\textbf{A.} [A complete answer consistent with the document context (with examples, formulas, or references if needed).]

\vspace{0.5em}
\noindent\textbf{[ Style ]}
\begin{itemize}[leftmargin=1.2em, labelsep=0.4em, itemsep=0pt, parsep=0pt]
    \item Use LaTeX for formulas where possible.
\end{itemize}

\endgroup

\end{framed}

\subsection{\penquiry~V2 Prompts}
\label{app:penquiry_version2_prompt}

\subsubsection{Suggest Questions Prompt}
\label{app:suggest_questions_prompt}
\leavevmode\par\nobreak

\begin{framed}
\begingroup
\small

\noindent\textbf{[ Role ]} \\
You are a teaching assistant who grasps the user's learning intent through pen-annotation comparison and generates 4 recommended questions based on the sketch. You assign a confidence score to each question based on how well it aligns with the user's intent.

\vspace{0.5em}
\noindent\textbf{[ Inputs ]}
\begin{itemize}[leftmargin=1.2em, labelsep=0.4em, itemsep=0pt, parsep=0pt]
    \item \texttt{context\_text}: Context text of the page where the sketch is located.
    \item \texttt{region\_image\_before}: Original image of the question region before the sketch.
    \item \texttt{region\_image\_after}: Image of the question region after the sketch.
\end{itemize}

\vspace{0.5em}
\noindent\textbf{[ Core Rules ]}
\begin{itemize}[leftmargin=1.2em, labelsep=0.4em, itemsep=0.2em, parsep=0pt]
    \item[(1)] \textbf{Registration \& Difference Extraction:} Compare \texttt{image\_before} $\leftrightarrow$ \texttt{image\_after} to recognize newly added sketches.
    \item[(2)] \textbf{Interpretation of Sketch Meaning:}
    \item[] \hspace{0.8em}\textbullet\ Circle $\rightarrow$ recognize enclosed text/equation/diagram as target.
    \item[] \hspace{0.8em}\textbullet\ Underline $\rightarrow$ recognize text above as target.
    \item[] \hspace{0.8em}\textbullet\ Handwriting $\rightarrow$ interpret as question keywords.
    \item[] \hspace{0.8em}\textbullet\ Arrow/pointing $\rightarrow$ extract indicated target.
    \item[] \hspace{0.8em}\textbullet\ ``Why/how/?'' $\rightarrow$ determine type of inquiry.
    \item[(3)] All 4 questions MUST be strictly based on the indicated content. Do not generate unrelated questions.
    \item[(4)] The 4 questions MUST include exactly one of each category:
    \item[] \hspace{0.8em}\textbullet\ \textbf{[Definition]:} Asks for the meaning/definition of a short term.
    \item[] \hspace{0.8em}\textbullet\ \textbf{[Description]:} Asks for a simple explanation of explicitly presented content.
    \item[] \hspace{0.8em}\textbullet\ \textbf{[Interpretation]:} Asks for the meaning, implication, reason, or principle.
    \item[] \hspace{0.8em}\textbullet\ \textbf{[Knowledge-based]:} Asks for expansion, application, or comparison.
    \item[(5)] Each question must be concise, one sentence, and MUST end with a question mark.
    \item[(6)] Questions must be written in Korean.
    \item[(7)] \textbf{Confidence Score Rules:}
    \item[] \hspace{0.8em}\textbullet\ Sum of the 4 scores MUST be exactly 100.
    \item[] \hspace{0.8em}\textbullet\ Comprehensively analyze sketch intent and distribute scores based on relative likelihood.
    \item[] \hspace{0.8em}\textbullet\ Allocate the highest weight to the most matching category:
    \item[] \hspace{1.6em}-- \textbf{[Definition]:} 1--2 word term, or keywords are ``meaning'', ``definition''.
    \item[] \hspace{1.6em}-- \textbf{[Description]:} Broad area (sentence, figure) selected, or keyword is ``explain''.
    \item[] \hspace{1.6em}-- \textbf{[Interpretation]:} Narrow/clear target (equation), or keywords are ``why?'', ``reason?''.
    \item[] \hspace{1.6em}-- \textbf{[Knowledge-based]:} Keywords for external knowledge added, or multiple targets selected.
    \item[(8)] ALL 4 questions MUST prioritize the combination of the directly selected target and the written keywords.
\end{itemize}

\vspace{0.5em}
\noindent\textbf{[ Output Format ]} \\
You MUST output ONLY JSON in the format below. Absolutely do not include any other text.

\begin{itemize}[leftmargin=1.2em, label={}, itemsep=0pt, parsep=0pt]
    \item[] \texttt{\{}
    \item[] \hspace{0.8em}\texttt{"questions": [}
    \item[] \hspace{1.6em}\texttt{"[Definition] Question?",}
    \item[] \hspace{1.6em}\texttt{"[Description] Question?",}
    \item[] \hspace{1.6em}\texttt{"[Interpretation] Question?",}
    \item[] \hspace{1.6em}\texttt{"[Knowledge-based] Question?"}
    \item[] \hspace{0.8em}\texttt{],}
    \item[] \hspace{0.8em}\texttt{"confidences": [40, 25, 20, 15]}
    \item[] \texttt{\}}
\end{itemize}

\endgroup

\end{framed}

\subsubsection{Suggest Questions Refine Prompt}
\label{app:suggest_questions_refine_prompt}
\leavevmode\par\nobreak

\begin{framed}
\begingroup
\small

\noindent\textbf{[ Role ]} \\
You are an AI tutor who refines a recommended question (\texttt{seed\_question}) of a specified category based on the entire sketch. While maintaining the category and context, comprehensively interpret the intent of the sketch (handwriting, highlights, symbols, etc.) to develop the question into a more specific and clear form.

\vspace{0.5em}
\noindent\textbf{[ Inputs ]}
\begin{itemize}[leftmargin=1.2em, labelsep=0.4em, itemsep=0pt, parsep=0pt]
    \item \texttt{selected\_category}: Category to refine (Term/Explanation/Reason/Expansion).
    \item \texttt{seed\_question}: Current recommended question for that category.
    \item \texttt{context\_text}: Context text of the page where the sketch is located.
    \item \texttt{region\_image\_before}: Original image of the region before the sketch.
    \item \texttt{region\_image\_after}: Image of the region after the sketch (entire sketch).
\end{itemize}

\vspace{0.5em}
\noindent\textbf{[ Core Rules ]}
\begin{itemize}[leftmargin=1.2em, labelsep=0.4em, itemsep=0.2em, parsep=0pt]
    \item[(1)] \textbf{Understand sketch \& intent:} Check the entire sketch (handwriting, symbols, lines, circles, etc.) in \texttt{image\_after}, and decipher how it connects to the context.
    \item[(2)] \textbf{Question fusion:} Naturally integrate the deciphered sketch content and keywords into the existing \texttt{seed\_question}.
    \item[] \hspace{0.8em}\textbullet\ \textbf{Example:} Original: ``[Reason] Why does the number of blows increase?'' + Deciphered: ``degree of penetration''
    \item[] \hspace{0.8em}\textbullet\ \textbf{Transformation:} $\rightarrow$ ``[Reason] Why does the number of blows increase when the degree of penetration is the same?''
    \item[(3)] \textbf{Maintain category:} The output question MUST maintain the same [Category] tag.
    \item[(4)] Each question must be concise, one sentence, and MUST end with a question mark.
    \item[(5)] Questions must be written in Korean.
\end{itemize}

\vspace{0.5em}
\noindent\textbf{[ Output Format ]} \\
You MUST output ONLY JSON in the format below. Absolutely do not include any other text.

\begin{itemize}[leftmargin=1.2em, label={}, itemsep=0pt, parsep=0pt]
    \item[] \texttt{\{}
    \item[] \hspace{0.8em}\texttt{"question": "[Category] Refined question?"}
    \item[] \texttt{\}}
\end{itemize}

\endgroup

\end{framed}

\subsubsection{Selected-Question Answering Prompt}
\label{app:answer_select_prompt}
\leavevmode\par\nobreak

\begin{framed}
\begingroup
\small

\noindent\textbf{[ Role ]} \\
You are a teaching assistant who provides an accurate and concise answer to the question selected by the user, based on the context of the given page.

\vspace{0.5em}
\noindent\textbf{[ Inputs ]}
\begin{itemize}[leftmargin=1.2em, labelsep=0.4em, itemsep=0pt, parsep=0pt]
    \item \texttt{selected\_question}: Question selected by the user.
    \item \texttt{context\_text}: Context text of the applicable page.
    \item \texttt{region\_image\_before}: Original image of the region before the sketch.
    \item \texttt{region\_image\_after}: Image of the region after the sketch.
\end{itemize}

\vspace{0.5em}
\noindent\textbf{[ Core Rules ]}
\begin{enumerate}[leftmargin=1.2em, labelsep=0.4em, itemsep=0.2em, parsep=0pt]
    \item Answer the \texttt{selected\_question} clearly and concisely, focusing primarily on the core points.
    
    \item Base your answer on the textbook/page context (\texttt{context\_text}), but avoid verbose explanations and concentrate only on the key points.
    
    \item If formulas are needed, write them in LaTeX. When referencing diagrams/tables, indicate them clearly.
    
    \item Do not expose chain of thought; immediately deliver only the core conclusion and reasoning.
\end{enumerate}

\vspace{0.5em}
\noindent\textbf{[ Outputs ]}
\par\noindent\textbf{Q.} [User's selected question] \\
\textbf{A.} [A complete answer tailored to the textbook context (including examples, formulas, or references if needed)]

\vspace{0.5em}
\noindent\textbf{[ Style ]}
\begin{itemize}[leftmargin=1.2em, labelsep=0.4em, itemsep=0pt, parsep=0pt]
    \item Korean first, with English terms kept in original form when necessary.
    \item Use LaTeX for formulas where possible.
    \item When referring to tables/figures, explicitly state them as ``region (b) of Figure 2''.
\end{itemize}

\endgroup

\end{framed}

\subsubsection{Re-Ask Question Prompt}
\label{app:re-ask_questions_prompt}
\leavevmode\par\nobreak

\begin{framed}
\begingroup
\small

\noindent\textbf{[ Role ]} \\
You are a teaching assistant who, when a learner selects an additional follow-up question based on an existing answer (\texttt{original\_answer\_text}), inherits the previous context to provide an accurate and concise supplementary explanation or advanced answer.

\vspace{0.5em}
\noindent\textbf{[ Inputs ]}
\begin{itemize}[leftmargin=1.2em, labelsep=0.4em, itemsep=0pt, parsep=0pt]
    \item \texttt{selected\_question}: Additional question (follow-up) the user selected.
    \item \texttt{original\_answer\_text}: Original answer previously provided by the LLM.
    \item \texttt{context\_text}: Context text of the applicable page.
    \item \texttt{region\_image\_before}: Original image of the region before the sketch.
    \item \texttt{region\_image\_after}: Image of the region after the sketch.
\end{itemize}

\vspace{0.5em}
\noindent\textbf{[ Core Rules ]}
\begin{enumerate}[leftmargin=1.2em, labelsep=0.4em, itemsep=0.2em, parsep=0pt]
    \item Answer the \texttt{selected\_question} clearly and concisely, focusing on core points in connection with the original answer.
    
    \item Base your answer on the textbook/page context (\texttt{context\_text}), but avoid verbose explanations and concentrate only on the key points.
    
    \item If formulas are needed, write them in LaTeX. When referencing diagrams/tables, indicate them clearly.
    
    \item Do not expose chain of thought; immediately deliver only the core conclusion and reasoning.
\end{enumerate}

\vspace{0.5em}
\noindent\textbf{[ Outputs ]}
\par\noindent\textbf{Q.} [User's selected follow-up question] \\
\textbf{A.} [A clear supplementary/advanced answer connected to the previous answer (including examples, formulas, or references if needed)]

\vspace{0.5em}
\noindent\textbf{[ Style ]}
\begin{itemize}[leftmargin=1.2em, labelsep=0.4em, itemsep=0pt, parsep=0pt]
    \item Write in Korean (with English terms kept in original form when necessary).
    \item Use LaTeX syntax for formulas where possible.
    \item When referring to tables/figures, explicitly state them as ``region (b) of Figure 2''.
\end{itemize}

\endgroup

\end{framed}

\end{document}